\RequirePackage{fix-cm}
\documentclass[smallextended]{svjour3}       
\smartqed  
\usepackage{graphicx}
\usepackage{natbib}
\setcitestyle{aysep={}}
\newcommand\citen[1]{\citeauthor*{#1}}						
\newcommand\citens[1]{\citeauthor*{#1}'s}					
\newcommand\cites[1]{\citeauthor*{#1}'s\ (\citeyear{#1})}	
\usepackage{xcolor}		
\usepackage{amssymb}	
\usepackage{amsmath}	
\usepackage{multirow}
\usepackage{tabularx}
\usepackage{enumitem} 	
\usepackage{pdflscape}
\usepackage{afterpage}
\begin{document}

\title{The impact of surplus sharing on the outcomes of specific investments under negotiated transfer pricing
}
\subtitle{An agent-based simulation with fuzzy Q-learning agents}

\titlerunning{Specific investments under negotiated transfer pricing}        

\author{Christian Mitsch
}


\institute{Christian Mitsch (0000-0002-8587-4887) \at
              Department of Management Control and Strategic Management, University of Klagenfurt \\
              Universit\"atsstra\ss{}e 65-67, 9020 Klagenfurt, Austria \\
              \email{christian.mitsch@aau.at}           
}

\date{}


\vspace*{3mm}
\begin{center}
 	\begin{Large}
 		\textbf{The impact of surplus sharing on the outcomes of specific investments under negotiated transfer pricing} \\
 	\end{Large}
 	\vspace{3mm}
 	\begin{large}
 		An agent-based simulation with fuzzy Q-learning agents \\
 	\end{large}
 	\vspace{10mm}
 	\begin{normalsize}
 		$\text{Christian Mitsch}^{[0000-0002-8587-4887]}$ \\
 		\vspace{3mm}
 		Department of Management Control and Strategic Management \\
 		University of Klagenfurt, 9020, Austria \\
 		\texttt{christian.mitsch@aau.at} 
              
 	\end{normalsize}
\end{center}

\vspace{10mm}

\begin{abstract}

	This paper focuses on specific investments under negotiated transfer pricing. 
	Reasons for transfer pricing studies are primarily to find conditions that maximize the firm's overall profit, especially in cases with bilateral trading problems with specific investments. 
	However, the transfer pricing problem has been developed in the context where managers are fully individual rational utility maximizers. 
	The underlying assumptions are rather heroic and, in particular, how managers process information under uncertainty, do not perfectly match with human decision-making behavior. 
	Therefore, this paper relaxes key assumptions and studies whether cognitively bounded agents achieve the same results as fully rational utility maximizers and, in particular, whether the recommendations on managerial-compensation arrangements and bargaining infrastructures are designed to maximize headquarters' profit in such a setting. 

	Based on an agent-based simulation with fuzzy Q-learning agents, it is shown that in case of symmetric marginal cost parameters, myopic fuzzy Q-learning agents invest only as much as in the classic hold-up problem, while non-myopic fuzzy Q-learning agents invest optimally.
	However, in scenarios with non-symmetric marginal cost parameters, a deviation from the previously recommended surplus sharing rules can lead to higher investment decisions and, thus, to an increase in the firm's overall profit. 







\vspace{5mm}

\keywords{Agent-based simulation \and Bounded rationality \and Fuzzy Q-learning \and Hold-up problem \and Negotiated transfer pricing \and Specific investments}
\end{abstract}

\newpage
\section{Introduction}
\label{sec:Introduction}

\subsection{Problem formulation}
\label{ss:Problem_formulation}

	As a business organization grows and the degree of the headquarters' task increases, delegating authority over operating decisions to managers might result in benefits, as managers, e.g., can use their superior knowledge to achieve better solutions for the firm \citep{wagenhofer1994transfer}. 
	In addition, the concept of profit centers can be applied to determine profit for internal divisions of responsibility and for purposes of management control, e.g., optimal resource allocation and profitability  \citep[e.g.,][]{kupper2013controlling, milgrom1992economics}. 
	Divisions' interests may be in conflict with the goal of the headquarters and, furthermore, divisions that are treated as profit centers often operate in a competitive environment within the firm, e.g., competing for financial resources. 
	Competitive behavior among profit centers of a firm may increase the profit of some divisions but does not necessarily maximize the headquarters' profit and, especially, they neglect the potentially negative consequences of their decisions for other divisions and the entire firm \citep{gox2006economic}. 

	The divisions' coordination problem increases when there is an internal transfer of goods and services from one division to another, e.g., transferring an intermediate product. 
	Internal transfers are often performed under conditions of asymmetric information \citep{baldenius2000intrafirm}. 
	For instance, one division has better knowledge about the costs of manufacturing the intermediate product, whereas another division has private information related to the net revenues from selling the intermediate product. 
	Therefore, transfer pricing as an instrument for coordinating operational decisions can be applied to manage potential conflicts and to guide the internal trade  \citep{baldenius1999negotiated}. 
	With a perfectly competitive market for the intermediate product, the optimal transfer price corresponds to the external market price, as this leads to efficient divisions' decision-making \citep{cook1955decentralization}. 
	If there is no external market for the intermediate product because the product is highly specialized, the headquarters could set the transfer price equal to the marginal cost of the intermediate product \citep{hirshleifer1956economics, schmalenbach1908verrechnungspreise}. 
	However, large business organizations usually do not have the opportunity, e.g., due to high costs or due to lack of information, to create an optimal production program of the entire firm centrally and, in generally, the divisions' cost information are required to optimally set the transfer price. 

	The headquarters' task is made more difficult when the divisions can additionally make specific investments that increase the value of internal trade. 
	The main problem with specific investments is that they are irreversible and are of little or no value in the divisions' external lines of business \citep{edlin1995specific}. 	
	For example, one division may incur an upfront fixed cost, e.g., an investment in production technology that reduces the variable costs of manufacturing intermediate products, whereas another division may improve the net revenues by investing in marketing activities. 
	However, due to the fact that the divisions bear the full cost of their own investments, but are evaluated according to their profits, each division is prone to underinvestment. 
	This problem is also known as the ``hold-up'' problem \citep{schmalenbach1908verrechnungspreise, williamson1979transaction, williamson1985economic}.
	
	The hold-up problem could be prevented by signing a complete contract that contains specific clauses for all possible future events that may occur after the investment decision  \citep{gox2006economic}. 
	But in reality, it is impossible to specify contract clauses for all anticipated eventualities. 
	For this reason, different transfer pricing mechanisms are designed which ensure the optimal solution of the headquarters' control problem. 
	
	The remainder of this paper is organized as follows: 
	The next section of the introduction covers a short literature review of negotiated transfer pricing models, followed by a criticism of key assumptions required to solve such models. 
	Then, a series of model relaxations to make assumptions about human decision-making behavior more realistic is given and the introduction concludes with the research method to study the hold-up problem with cognitively bounded agents. 
	In Sec. \ref{sec:The_Model}, negotiated transfer pricing for bilateral negotiation problems with specific investments is introduced and the resulting solutions are discussed. 
	Sec. \ref{sec:agent_based_simulation} introduces the relaxed assumptions and describes what the agents can and cannot do in the agentized version of the negotiated transfer pricing model.\footnote{	
	``Agentized'' is derived from agentization and describes the process of turning a neoclassical model into an agent-based model \citep[e.g.,][]{guerrero2011using, leitner2015efficiency}. %
	}
	In addition, Sec. \ref{sec:agent_based_simulation} formalizes how fuzzy Q-learning works and gives a brief introduction to the exploration policies which are used in the simulation study. 
	The parameter settings for the agent-based simulation are explained in Sec. \ref{sec:Parameter_Settings_and_Simulation_Setup} and, in Sec. \ref{sec:Results_and_Discussion}, the results of the agent-based simulation are presented and discussed. 
	Sec. \ref{sec:Conclusion} contains concluding remarks and suggests possible directions for future research. 
	
	Please be aware that, in the following, the additional model designation ``specific investment'' is sometimes omitted for the sake of brevity. 

\subsection{Brief literature review}
\label{ss:Literature_review}	

	The economic transfer pricing research is typically divided into market-based, cost-based, and negotiated transfer pricing methods, which are usually based on game-theoretic models.\footnote{
	For a survey of the transfer pricing literature see, e.g., \cite{gox2006economic}, \cite{wagenhofer1994transfer}, and, for a survey in German see, e.g., \cite{ewert2014interne}, \cite{pfaff2004verrechnungspreise}, \cite{wagner2008corporate}. Readers unfamiliar with agency theory and game theory should refer to \cite{laffont2009theory} and \cite{osborne1994course}, respectively.} 
	Since it is generally impossible to give a recommendation for the ``best'' transfer pricing method, a variety of solutions to the transfer pricing problem are offered  \citep{gox2006economic}. 
	This paper focuses on negotiated transfer pricing, as is a common way of setting the transfer price in multidivisional firms \citep[e.g.,][]{cook1955decentralization, eccles1985transfer}.\footnote{
	In a negotiated transfer pricing model, divisions' decisions are free from headquarters interventions. Such models stand in contrast to administered transfer pricing models \cite[][]{eccles1988price}.}

	The starting point of the research is the well-known negotiated transfer pricing model by \cite{edlin1995specific}. 	
	The authors extend the neoclassical model of \cite{schmalenbach1908verrechnungspreise} and \cite{williamson1985economic} by assuming that each division can simultaneously make a specific investment that enhances the value of internal trade before the negotiation over transfer price and transfer quantity takes place. 
	\cite{edlin1995specific} show that efficient investments can be guaranteed, if (1) an upfront fixed-price contract which determines the quantity of the intermediate product to be transferred and the corresponding transfer price and (2) the opportunity to renegotiate this contract once cost and revenue information are available to both divisions. 
	In particular, the transfer pricing scheme by \citen{edlin1995specific} seems to be robust in the sense that, for any combination of divisions' bargaining power, the hold-up problem can be solved. 
	It should be noted that their results depend on the main assumption that both divisions have symmetric information about costs and revenues in their bilateral negotiation. 
	
	In contrast to an upfront fixed-price contract between divisions, \cite{pfeiffer2004value} adds an additional phase to the sequence of events within the negotiated transfer pricing model in which the headquarters may grant individual divisions access to ``pre-investment'' information. 
	This phase should help the divisions to revise their beliefs about the realizations of costs and revenues, before each division makes an investment decision independently of each other. 
	\citen{pfeiffer2004value} shows that installing a pre-investment information system has an ambiguous effect on the undertaken investments. 
	On one hand, it can mitigate the hold-up problem and even lead to efficient investments, but it can also cause dysfunctionalities for the entire firm. 
	For a perfect pre-investment information system, \citen{pfeiffer2004value} provides conditions for the class of linear surplus sharing rules under which the headquarters' profit is maximized. 
	
	Finally, well-known contributions of negotiated transfer pricing under conditions of asymmetric information are given, e.g, by \cite{ronen1988approach} or by \cite{vaysman1998model}. 
	For instance, \citen{vaysman1998model} considers a negotiated transfer pricing model in a multidivisional firm where each division has better divisional information than the headquarters and the other division. 
	\citens{vaysman1998model} analysis indicates that in scenarios with information asymmetry, managerial-compensation arrangements and bargaining infrastructures can be designed to maximize headquarters' profit. 

\subsection{Criticism of model assumptions}
\label{ss:Criticism_model_assumptions}	

	However, the transfer pricing problem has been developed in the context where divisions are fully individual rational utility maximizers. 
	For instance, \cite{wagenhofer1994transfer} uses backward induction to compute the subgame perfect equilibrium of his modified negotiated transfer pricing model, or \cite{vaysman1998model} applies the concept of perfect Bayesian equilibrium, developed by \cite{fudenberg1991perfect}, to extend the idea of subgame perfection to games of incomplete information. 
	The concept of subgame perfect equilibrium is attractive, but it should be mentioned that it has been sometimes criticized because it depends excessively on rationality \citep[e.g.,][]{aumann1989game, laffont2009theory}. 
	For example, how could a division in its investment decision assume that the other division will behave optimally at the later negotiation as needed by subgame perfect implementation? 
	Such concepts require heroic assumptions about the information the divisions have regarding, for example, the structure of the whole game, the underlying utility functions, and the knowledge about opponent's behavior \citep{sebenius1992negotiation}, or certain conditions about the probability distributions of their environment \citep{simon1979rational}. 
	In practice, these assumptions are often not met, but, to put it in the words of \cite{aumann1989game}, ``The common knowledge assumption underlies all of game theory and much of economic theory.''. 
	
	Apart from this, one could also criticize that the expectations of how divisions interact with each other in a multi-stage decision-making process and, in particular, how they process information under uncertainty, do not perfectly match with human decision-making behavior. 
	The relevant literature also suggests that the assumption of rational behavior of economic agents, as it is common in neoclassical economics, is too strong \citep[e.g.,][]{arthur1991designing, richiardi2006common, simon1955behavioral}. 
	In particular, assumptions on rationality are implausible in models that reflect human behavior and, therefore, most of such models quickly become impractical and unrealistic \citep{young2001individual}. 

\subsection{Relaxation of assumptions and research method}
\label{ss:Relaxation_key_assumptions}		
	Against this background, an agent-based variant of negotiated transfer pricing is analyzed in which (1) divisions have no beliefs about the rationality of other divisions nor how other divisions face their maximization problem, (2) they have no common knowledge of the opponent's utility function, and (3) they are subject to cognitive limitations such as limits on the capacity of human memory, restrictions on the speed of learning, and constraints in foresight. 
	To deal with the divisions' bounded rationality, asymmetric information, and cognitive limitations, a computer simulation with individual learning agents is set up for the before-mentioned hold-up problem. 
	
	The learning method in this paper rests on reinforcement learning because using reinforcement learning in an agent-based simulation may be more realistic as reinforcement learning provides methods that are close to the way animals and humans learn things \citep[e.g.,][]{collins2019reinforcement, niv2009reinforcement, sutton2018reinforcement}. 
	In particular, fuzzy Q-learning, which is ``technically'' based on dynamic programming and Monte Carlo methods, is used in this simulation study to describe the before-mentioned cognitively bounded agents.\footnote{ 
	The combination of dynamic programming and Monte Carlo ideas is called temporal-difference learning, a class of algorithms in reinforcement learning. 
	Readers unfamiliar with reinforcement learning methods should refer to \cite{sutton2018reinforcement}. 
	For agent-based simulations with Q-learning agents acting in an economic context see, e.g., \cite{dearden1998bayesian, tesauro2002pricing, waltman2008learning} and with fuzzy Q-learning agents see, e.g., \cite{kofinas2018fuzzy, rahimiyan2006modeling}.} 
	
	Since the model investigated here is characterized by a high degree of heterogeneity with respect to the structure and the dynamic interactions between agents and their environment, agent-based simulation seems to be an appropriate research method to deal with the complexity of the research problem. 
	In addition, an agent-based simulation also allows to observe the agents' behavior as well as the system's behavior on the macro-level in time, which otherwise cannot be derived in relation to a ``functional relationship'' from the individual behaviors of those agents \citep[e.g.,][]{epstein2006generative, wall2016agent}. 
	
	This paper enriches the research on negotiated transfer pricing by an agent-based simulation with fuzzy Q-learning agents. 
	In particular, this simulation study aims to provide a contribution to the hold-up problem by answering whether cognitively bounded agents achieve the same results as fully rational utility maximizers and, especially, whether the recommendations on managerial-compensation arrangements and bargaining infrastructures maximize headquarters' profits under more realistic assumptions. 

\section{Negotiated transfer pricing with fully rational agents}
\label{sec:The_Model}

\subsection{Organizational setting}
\label{ssec:Organizational_Setting}

\begin{figure}[h!]
		\centering
		\includegraphics[width=0.9\textwidth]{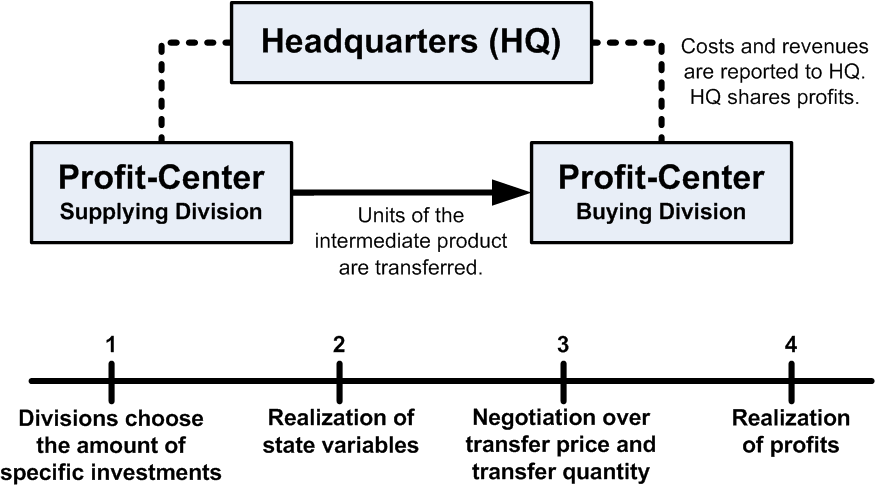}
		\caption{Schematic representation of the decentralized firm. In addition, the sequence of events within the negotiated transfer pricing model is illustrated. Source: Based on \cite{wagner2008corporate}.}
		\label{fig:Schematic_Representation}
	\end{figure}

	This paper analyzes a decentralized firm consisting of one headquarters and two divisions - the supplying division (the ``upstream'' division) and the buying division (the ``downstream'' division). 
	In line with the previous literature \cite[e.g.,][]{eccles1988price, edlin1995specific, gox2006economic, pfeiffer2007rekonstruktion, vaysman1998model, wagner2008corporate}, it is assumed that both divisions are organized as profit centers, i.e., each division is evaluated on its own divisional profit, and that the headquarters delegates to its divisions operating decisions like the amount of specific investment, the level of the transfer price, and the amount of intermediate products. 
	Suppose that the supplying division purchases raw materials from a commodity market in order to manufacture an intermediate product which is transferred to the buying division.\footnote{
	In negotiated transfer pricing models, it is commonly assumed that the intermediate product is highly specialized and, therefore, there is no external market for the intermediate product \citep[e.g.,][]{anctil1999negotiated, edlin1995specific, wagner2008corporate}. For the sake of simplicity, it is assumed that one unit of raw material is required to manufacture one unit of the intermediate product.} 
	On the other side, the buying division refines the intermediate product, e.g., to improve its quality, and, finally, sells it on an outlet market. 
	Figure \ref{fig:Schematic_Representation} schematically represents the negotiated transfer pricing model investigated here. 
	
	In the following, supplying division, buying division, and headquarters are abbreviated to $S$, $B$, and $HQ$, respectively. 
	The supplying division's costs of manufacturing $q \in \mathbb{R}^+$ units of the intermediate product are given by 
	\begin{equation}
		C_S(q,\theta_S,I_S) = (\theta_S - I_S ) \; q \;, \label{eq:supplying_division_costs}	
	\end{equation}
	where $\theta_S \in \mathbb{R}^+$ is a state variable which reflects the purchase price of raw materials and $I_S \in \mathbb{R}^+$ stands for the amount of specific investment carried by the supplying division. 
	In contrast, the buying division's net revenue is 
	\begin{equation}
		R_B(q,\theta_B,I_B) = (\theta_B - \frac{1}{2} \; b \; q + I_B) \; q \; , \label{eq:buying_division_net_revenue}
	\end{equation}  
	where $\theta_B \in \mathbb{R}^+$ is another state variable which represents the constant term in the inverse demand function for the selling product, $b \in \mathbb{R}^+$ describes the slope of the inverse demand function, and $I_B \in \mathbb{R}^+$ denotes the amount of specific investment carried by the buying division. 
	To keep the analysis simple and, especially, that the negotiated transfer pricing model has a unique subgame perfect equilibrium, it is assumed that the inverse demand function and the costs of production are linear and, for linguistic reasons, the term ``investment decision'' or just ``investment'' is used for $I_j$, $j \in  \{ S,B \}$, instead of ``amount of specific investment''. 
	
	It is assumed that the supplying division and the buying division have private information about the commodity market and the outlet market, respectively, which are embodied in the state variables $\theta_S$ and $\theta_B$. 
	In addition, supplying and buying division have information about the expected value of the outlet market and commodity market, respectively. 
	For the sake of simplicity, the state variables $\theta_S$ and $\theta_B$ are stochastically independent random variables.
	Further, the analysis assumes that the headquarters cannot observe the state variables nor the undertaken investments; 
	the headquarters only knows the expected values of the state variables.
	Furthermore, the headquarters' accounting system receives costs and revenues after the negotiation phase and, subsequently, the headquarters' profit $\Pi_{HQ}$ is calculated. 
	\begin{equation}
		\Pi_{HQ}(q,\theta_S,\theta_B,I_S,I_B) = R_B(q,\theta_B,I_B) - C_S(q,\theta_S,I_S) - w_S(I_S) - w_B(I_B)  \label{eq:headquarters_profit}
	\end{equation}
	Both the supplying division's investments and the buying division's investments cause divisional investment costs (or divisional capital expenditure) $w_S(I_S)$ and $w_B(I_B)$, respectively, which have a quadratic cost structure 
	\begin{equation}
		w_j(I_j) = \frac{1}{2} \; \lambda_j \; I_j^2 \; \; \; \text{for } j \in \{ S,B\} \;, \label{eq:divisional_investment_costs}
	\end{equation}
	where $\lambda_j \in \mathbb{R}^+$ denotes the marginal cost parameter. 
	It is assumed that the parameters $w_S$, $w_B$, $\lambda_S$, $\lambda_B$, and $b$ are known to the divisions and the headquarters. 
	Incidentally, the term, one half, is only for reasons of expediency. 
	Please notice that the supplying division's net revenue generated by the internal quantity transfer corresponds to the buying division's transfer costs, i.e., $R_S=C_B=TP \cdot q$, and, hence, they do not appear in the headquarters' profit $\Pi_{HQ}$. 
	In of almost all transfer pricing models, the formulas are represented by the transfer quantity $q$; the transfer price $TP$ does not appear in any formula, but $TP$ can be calculated explicitly by using substitution. 	
	
	In a nutshell, the negotiated transfer pricing model investigated here involves four dates (see Fig. \ref{fig:Schematic_Representation}): 
	At date one, both divisions have to make an investment decision independently of each other. 
	Subsequently, supplying and buying division independently observe the state variables $\theta_S$ and $\theta_B$, respectively. 
	At date three, the amount of intermediate products is determined by the divisions and, finally, the headquarters' profit as well as the divisions' profits are realized. 

	\subsection{First-best solution}
	\label{ss:First_Best_Solution}

		From the view of the headquarters, the decisions over investments and quantity are taken in such a way that Eq. \ref{eq:headquarters_profit} is maximized. 
		The solution of the two-stage decision problem is received by backward induction starting with the quantity decision on date three.\footnote{ 
		The solutions presented here, as well as the solutions of the most common transfer pricing methods, are described in \cites{wagner2008corporate} PhD thesis in more detail.} 
		\begin{equation}
			q^* \in \underset{q \in \mathbb{R}^+}{\mathrm{arg\,max}} \; \Pi_{HQ}(q,\theta_S,\theta_B,I_S,I_B) 
		\end{equation}
		Note, first-best solutions are indexed by a superscript $^*$ 
		and, in order to guarantee that the first-best solutions are reached, divisions must be fully rational at each stage of the decision problem, i.e., at each stage, the action with the highest outcome must be selected. 
		Since each division has access to all relevant information, backward induction can be applied to calculate the subgame perfect equilibrium. 
		
		On date three, the first order condition for maximizing the headquarters' profit $\Pi_{HQ}$ with respect to $q$ is equal to 
		\begin{equation}
			\frac{\partial \Pi_{HQ}}{\partial q} = \theta_B - b \; q + I_B - \theta_S + I_S = 0 
		\end{equation}
		and, hence, the profit maximizing quantity is
		\begin{equation}
			q^* = \frac{\theta_B - \theta_S + I_S + I_B}{b} \; . \label{eq:backward_q}
		\end{equation}
		On date one, the expected headquarters' profit $E[\Pi_{HQ}]$ is maximized with respect to $I_j$, $j \in \{ S,B \}$, because investment decisions are made under uncertainty. 
		\begin{equation}
			I_j^* \in \underset{I_j \in \mathbb{R}^+}{\mathrm{arg\,max}} \; E[\Pi_{HQ}(q,\theta_S,\theta_B,I_S,I_B)] 	\label{eq:backward_I}
		\end{equation}
		Differentiating the expected headquarters' profit with respect to investments yields 
		\begin{equation}
			I_j^* = \frac{E[q^*(\theta_S,\theta_B,I_S,I_B)]}{\lambda_j} \; . \label{eq:optimal_I_j}
		\end{equation}
		Now, substituting Eq. \ref{eq:backward_q} into Eq. \ref{eq:backward_I}, one gets the following first-best expected investments. 
		\begin{eqnarray}
			I_S^* & = & \frac{\lambda_S \; E[\theta_B - \theta_S]}{b \; \lambda_S \; \lambda_B - \lambda_S - \lambda_B} \label{eq:optimal_I_S} \\
			I_B^* & = & \frac{\lambda_B \; E[\theta_B - \theta_S]}{b \; \lambda_S \; \lambda_B - \lambda_S - \lambda_B} \label{eq:optimal_I_B}
		\end{eqnarray}
		Finally, substituting Eq. \ref{eq:optimal_I_S} and Eq. \ref{eq:optimal_I_B} into Eq. \ref{eq:backward_q} implies the first-best expected quantity. 
		\begin{equation}
			q^* = \frac{\theta_B - \theta_S}{b} + \frac{(\lambda_S + \lambda_B) \; E[\theta_B - \theta_S]}{b \; (b \; \lambda_S \; \lambda_B - \lambda_S - \lambda_B)} 	\label{eq:optimal_q}
		\end{equation}
		With the first-best solutions of the two-stage decision problem, the first-best expected profit is given by the following expression. 
		\begin{equation}
			\Pi_{HQ}^* = \frac{1}{2} \; \frac{\lambda_S \; \lambda_B \; E[\theta_B - \theta_S]^2}{b \; \lambda_S \; \lambda_B - \lambda_S - \lambda_B} + \frac{Var[\theta_B - \theta_S]}{2b} \label{eq:optimal_Profit_HQ}
		\end{equation}
		The headquarters' profit in Eq. \ref{eq:optimal_Profit_HQ} can be seen as a benchmark for the highest feasible profit that can be achieved, if investments and quantity are set equal to Eq. \ref{eq:optimal_I_S}, Eq. \ref{eq:optimal_I_B}, and Eq. \ref{eq:optimal_q}, respectively. 

	\subsection{Second-best solution}
	\label{sec:Second_Best_Solution}

		In negotiated transfer pricing models, both divisions simultaneously undertake investments before the quantity negotiation takes place. 
		Due to the fact that each division is free to decide any amount of investment and quantity, the headquarters has to provide an incentive compatibility for each division. 
		In order to ensure that both divisions report their state variables truthfully and the negotiation leads to an ex post efficient quantity, the following division incentives are commonly used.\footnote{
		In general, incentives for efficient investments do not necessarily provide incentives for an efficient quantity and verse versa. For instance, in full-cost transfer pricing models, both divisions are rewarded for efficient investments, but the negotiation does not lead to an ex post efficient quantity \citep[][]{baldenius1999negotiated}.} 
		\begin{eqnarray}
			\Pi_S(q,\theta_S,\theta_B,I_S,I_B,\Gamma) & = & \Gamma \; M(q,\theta_S,\theta_B,I_S,I_B) - w_S(I_S) \label{eq:Profit_S} \\
			\Pi_B(q,\theta_S,\theta_B,I_S,I_B,\Gamma) & = & (1-\Gamma) \; M(q,\theta_S,\theta_B,I_S,I_B) - w_B(I_B) \label{eq:Profit_B} 
		\end{eqnarray}	
		$\Pi_S$ and $\Pi_B$ denote the profit of the supplying division and the profit of the buying division, respectively. 
		$M(q,\theta_S,\theta_B,I_S,I_B)$ stands for the headquarters' contribution margin, i.e., $M=R_B-C_S$, and $\Gamma \in [0,1]$ represents the share of the contribution margin achieved. 
		In almost all negotiated transfer pricing models, the negotiation process is modeled by a simple linear surplus sharing rule, i.e., the supplying division receives a share $\Gamma$ and the buying division a fraction $(1-\Gamma)$ of the headquarters' contribution margin. 
		The parameter $\Gamma$ can be considered as the supplying division's bargaining power and, for $\Gamma=0.5$, the negotiation result is equivalent to the Nash bargaining solution of the standard Rubinstein bargaining game \citep[e.g.,][]{baldenius1999negotiated, rubinstein1982perfect}. 

		Again, the analysis of negotiated transfer pricing starts by backward induction on date three. 
		Since the headquarters delegates operation decisions to its divisions, the supplying division and the buying division seek to maximize their own divisional profits which are given by Eq. \ref{eq:Profit_S} and Eq. \ref{eq:Profit_B}, respectively. 
		Starting with the quantity decision on date three, one gets
		\begin{equation}
			q^{sb} = \frac{\theta_B - \theta_S + I_S + I_B}{b} \; . \label{eq:backward_q_sb}
		\end{equation}
		A closer look reveals that differentiating the divisional profits with respect to $q$ leads to the same profit maximizing quantity given by Eq. \ref{eq:backward_q}, but the first order conditions for maximizing the expected divisional profits $E[\Pi_j]$, $j \in \{ S,B \}$, with respect to $I_j$ lead to second-best investments (labelled by the superscript $sb$). 
		\begin{eqnarray}
			I_S^{sb} & = & \frac{\Gamma \; E[q^{sb}(\theta_S,\theta_B,I_S,I_B)]}{\lambda_S} \label{eq:second_best_I_j} \\
			I_B^{sb} & = & \frac{(1-\Gamma) \; E[q^{sb}(\theta_S,\theta_B,I_S,I_B)]}{\lambda_B}
		\end{eqnarray}
	
		In the case of negotiated transfer pricing, the headquarters has to determine how the contribution margin is shared between the divisions. 
		Therefore, the headquarters is faced with the following decision problem which has to be solved before the decision-making process for the divisions begins. 
		\begin{equation}
			\Gamma^{sb} \in \underset{\Gamma \in [0,1]}{\mathrm{arg\,max}} \; E[\Pi_{HQ}(q,\theta_S,\theta_B,I_S,I_B,\Gamma)] 
		\end{equation}
		Differentiating the expected headquarters' profit with respect to $\Gamma$ 
		\begin{equation}
			\frac{\partial}{\partial \Gamma} \Big( \big( E[\theta_B]-\frac{1}{2} b q^{sb} + I_B^{sb} \big) q^{sb} - \big( E[\theta_S] - I_S^{sb} \big) q^{sb} -\frac{1}{2} \lambda_S {I_S^{sb}}^2 -\frac{1}{2} \lambda_B {I_B^{sb}}^2 \Big) = 0
		\end{equation}
		yields
		\begin{equation}
			\Gamma^{sb} = \frac{b \; \lambda_B - 1}{b \; (\lambda_S + \lambda_B) - 2} \; . \label{eq:Gamma_sb}
		\end{equation}
		By using substitution as in Sec. \ref{ss:First_Best_Solution}, one obtains the following second-best solutions.  
		\begin{eqnarray}
			I_S^{sb} & = & \frac{\lambda_B \; E[\theta_B - \theta_S]}{(b \; \lambda_S -1) \; (\lambda_S + \lambda_B)} \label{eq:second_best_I_S} \\
			I_B^{sb} & = & \frac{\lambda_S \; E[\theta_B - \theta_S]}{(b \; \lambda_B -1) \; (\lambda_S + \lambda_B)} \label{eq:second_best_I_B} \\
			q^{sb} & = & \frac{\lambda_S \; \lambda_B \; (b \; (\lambda_S + \lambda_B)-2) \; E[\theta_B - \theta_S]}{(b \; \lambda_S -1) \; (b \; \lambda_B -1) \; (\lambda_S + \lambda_B)} + \frac{\theta_B - \theta_S - E[\theta_B - \theta_S]}{b} \label{eq:second_best_q} \\
			\Pi_{HQ}^{sb} & = & \frac{1}{2} \; \frac{\lambda_S \; \lambda_B \; (b \; ( \lambda_S + \lambda_B)-1) \; E[\theta_B - \theta_S]^2}{(b \; \lambda_S -1) \; (b \; \lambda_B -1) \; (\lambda_S + \lambda_B)} + \frac{Var[\theta_B - \theta_S]}{2b} \label{eq:second_best_HQ}	
		\end{eqnarray}
		According to Eq. \ref{eq:Gamma_sb}, the headquarters assigns higher bargaining power to the division that has a lower marginal cost parameter. 
		If $\lambda_S=\lambda_B$, then the surplus parameter is one half, i.e., both divisions receive equal shares of the contribution margin achieved. 
		Regardless of the headquarters' decision on $\Gamma$, Eq. \ref{eq:second_best_I_S} and Eq. \ref{eq:second_best_I_B} indicate that both divisions are prone to underinvesting and, hence, the first-best quantity resulting from the negotiation process cannot be reached either. 
		As outlined in the first-best case, the headquarters' profit depends not only on $\lambda_j$, $j \in \{ S,B \}$, but also on the expected values of the state variables and benefits, in particular, from the volatility of the markets.\footnote{
		Please be aware that the variance of the difference between two stochastically independent random variables is given by the sum of their variances. In general it holds: $Var[\theta_B \pm \theta_S] = Var[\theta_B] + Var[\theta_S] \pm 2Cov[\theta_B,\theta_S]$.} 











	\vspace{20mm}

\section{An agent-based variant of negotiated transfer pricing with cognitively bounded agents}
\label{sec:agent_based_simulation} 

	\subsection{Relaxed assumptions}
	\label{ssec:Relaxed_Assumptions}

		Recall that the negotiated transfer pricing model \citep[e.g.,][]{edlin1995specific, fudenberg1991perfect, vaysman1998model} requires fully individual rational utility maximizers. 
		As mentioned in the introduction, such assumptions are rather heroic. 
		Therefore, this paper investigates negotiated transfer pricing with divisions which (1) have no beliefs about the rationality of other divisions nor how other divisions face their maximization problem, (2) have no common knowledge of the opponent's utility function, and (3) are subject to cognitive limitations such as limits on the capacity of human memory, restrictions on the speed of learning, and constraints in foresight. 
		In order to implement these relaxations of key assumptions, an agent-based simulation is set up. 
		
		The negotiated transfer pricing model presented in the previous section assumes that the divisions make decisions by means of their anticipated information. 
		Hence, the common knowledge assumption is required to determine the optimal values of the decision variables $\Gamma$, $I_j$, and $q$, $j \in \{ S,B\}$. 
		In the agentized version of the negotiated transfer pricing model, the divisions know neither the profit function $\Pi_j$ nor the expected value of the state variable $E[\theta_j]$ of the other division. 
		In addition, both divisions only share those parameters that are necessary for an efficient quantity decision in their negotiation phase at date three. 
		Table \ref{tab:Overview} shows the parameters of negotiated transfer pricing presented in Sec. \ref{sec:The_Model} and summarizes the most important changes associated with the agent-based variant of negotiated transfer pricing with cognitively bounded agents.
		It should be emphasized once again that, in this paper, cognitively bounded agents are modeled by fuzzy Q-learning agents, which are introduced in Sec. \ref{sec:The_learning_method}. 
		
		\afterpage{
		\begin{landscape}
		\begin{table}[h!]
		\linespread{1.1}\selectfont
		\centering
		\caption{Comparison between negotiated transfer pricing with fully rational agents and the agent-based variant with cognitively bounded agents.}
		\label{tab:Overview}
		\begin{tabularx}{186mm}{|c|l|l|l|}
			\hline
			\multirow{2}{*}{Parameter} & \multirow{2}{*}{Description} & Negotiated transfer pricing  & Agent-based variant of negotiated transfer pricing \\
			& & with fully rational agents (see Sec. \ref{sec:The_Model}) & with cognitively bounded agents (see Sec. \ref{sec:agent_based_simulation}) \\
			\hline
			\rule{0pt}{9pt} $\Pi_{HQ}$ & headquarters' profit & common knowledge & common knowledge \\
			\rule{0pt}{12pt} $\Pi_j$ & division's profit & common knowledge & private information for entire period \\
			\rule{0pt}{12pt} $E[\theta_j]$ & expected value of state variable & common knowledge & private information for entire period \\
			\rule{0pt}{12pt} $C_j$ & division's costs & common knowledge & private information until negotiation \\
			\rule{0pt}{12pt} $R_j$ & division's net revenue & common knowledge & private information until negotiation \\
			\rule{0pt}{12pt} $\lambda_j$ & division's marginal cost & common knowledge & private information until negotiation \\
			\rule{0pt}{12pt} $w_j$ & division's investment costs & common knowledge & private information until negotiation \\
			\rule{0pt}{12pt} $b$ & slope of the inv. demand func. & common knowledge & private information until negotiation \\
			\hline
			\rule{0pt}{9pt} $\Gamma$ & surplus sharing parameter & is set optimally & is a scenario-based exogenous parameter, which \\
			& & & varies in small steps \\
			\rule{0pt}{12pt} $I_j$ & amount of specific investment & is set optimally given $\Gamma$ & is chosen by an exploration policy, which mainly \\
			& & & depends on the learned Q-function \\
			\rule{0pt}{12pt} $\theta_j$ & state variable & private information until negotiation & private information until negotiation \\
			\rule{0pt}{12pt} $q$ & quantity & is set optimally given $I_j$ and $\theta_j$ & is set optimally given $I_j$ and $\theta_j$ \\
			\hline
		\end{tabularx}
		\linespread{1}\selectfont
		\end{table}
		\end{landscape}
		}
		
		\subsection{What the agents have to learn}		
		
		In the agent-based variant, both divisions have first to learn how their environment works in which they are and, in order to do so, they have the ability to store their individually learned information. 
		For this purpose, the divisions are able to process a manageable number of states (concretely, $25$ states are used in the simulation study to describe the agent's memory; the parameter settings for the simulations are discussed in Sec. \ref{sec:Parameter_Settings_and_Simulation_Setup} in more detail). 
		Moreover, the divisions have a learning rate which indicates how much new information overrides old information and both divisions have a discount factor which represents the agent's foresight, i.e., how important future rewards are to the agent. 
		While the learning rate is fixed in the simulation study, the discount factor is a scenario-based exogenous parameter which varies in small steps. 
		In addition to varying the discount factor, three different exploration policies are examined, as the agents no longer instantaneously see the consequences of their actions. 
		
		Recall that, in the organizational setting investigated here, the headquarters determines the incentives for the divisions in such a way that the negotiation over transfer price and quantity at date three leads to an ex post efficient quantity (cf. Eq. \ref{eq:backward_q_sb}). 
		Consequently, the two divisions are only faced with the investment decision at date one. 
		Therefore, the agents have to learn which amount of investment leads to which consequence. 
		Since the simulation study assumes that the agents have no prior knowledge about their consequences nor about opponent's decision-making behavior, the sequence of events within the negotiated transfer pricing model is run through several times. 
		In order to indicate this additional time dimension, a time subscript $t \in \mathbb{N}$ for the ``inner loop'' of the simulation is added to the model (a flow diagram of the agent-based simulation is given in Sec. \ref{sec:Parameter_Settings_and_Simulation_Setup}, Fig. \ref{fig:Procedure}). 
		

\subsection{The learning method}
\label{sec:The_learning_method}

	A common approach to describe individual learning behavior is reinforcement learning \citep{sandholm1996multiagent}. 
	This approach is applicable in cases where the learning individuals (agents) are in an unknown environment and choose actions for which they receive rewards, but they do not have the ability to directly observe their opponents' actions and, especially, they do not have beliefs about the likely play of others \citep{feltovich2000reinforcement}. 
	In reinforcement learning, a learning agent is able to perceive its environment, can choose actions that can change its environmental state, and, of course, the agent pursues a goal. 
	Moreover, an agent uses its past experiences to find an optimal policy to maximize its utility function.\footnote{ 
	Note, a policy is a mapping from environmental states to agent's actions and determines the learning agent's way of behaving \citep{sutton2018reinforcement}.}
	In the following, a brief introduction to Q-learning and its fuzzy version is provided. 

		\subsubsection{Q-learning}
		\label{sssec:Q_Learning_Agents}
	
			Q-learning is a popular reinforcement learning approach to multi-agent problems. 
			The learning method by \cite{watkins1989learning} is based on estimating the state-action value function (or Q-function) that ultimately gives the expected utility of a given action in a given state.
			In Q-learning, an agent learns its Q-values $\in \mathbb{R}$ by a simple iteration update which is given by
			\begin{equation}
				Q_{t+1}[\pmb{s}_t,a_{t}] = (1-\alpha) \; Q_{t}[\pmb{s}_t,a_{t}] + \alpha \; \Big( r_t(\pmb{s}_t,a_t) + \gamma \; \max\limits_{a \in \mathcal{A}} Q_t[\pmb{s}_{t+1},a] \Big) \; , \label{eq:Q_Update_Rule} 	
			\end{equation}
			where $\pmb{s}_t = (s_{t,S},s_{t,B}) \in \mathcal{S} \subset \mathbb{N}^2, a_t \in \mathcal{A} \subset \mathbb{N}$, and $r_t \in \mathbb{R}$ stand for state vector, action, and reward  at time $t \in \mathbb{N}$, respectively. 
			For the sake of readability, the subscript $j \in \{ S,B\}$ is omitted from all variables in the formulas. 
			In line with the negotiated transfer pricing model in Sec. \ref{sec:The_Model}, action $a_t$ taken by the supplying division (buying division) corresponds to the investment decision $I_S$ ($I_B$), the state $s_{t,S}$ ($s_{t,B}$) represents the last action $a_{t-1}$ chosen by the supplying division (buying division), and reward $r_t$ that the supplying division (buying division) receives is equal to $\Pi_S$ ($\Pi_B$). 
			
			$\mathcal{S}$ and $\mathcal{A}$ describe the state space and the action space, respectively. 
			Further, $\gamma \in [0,1)$ describes the discount factor and $\alpha \in (0,1]$ is the learning rate.\footnote{
			Technical note: If the discount factor is one, then the sum of the expected future rewards becomes infinite, which implies that the Q-values keep growing as long as the simulation is running. Hence, the Q-function does not converge.} 
			Q-learning can be seen as a weighted average of old stored information and new received information from the agent's environment. 
			Note that the values of the state-action value function are stored in a lookup table (labelled with square brackets) initialized with zero for all states and actions, i.e., there is no information about the agent's environment when the simulation is started. 

		\subsubsection{Fuzzy Q-learning}
		\label{sssec:Fuzzy_Q_Learning_Agents}		
			
			In the conventional Q-learning approach, the Q-values are recorded in a look-up table. 
			If the number of states and actions increases, this method is highly impractical and, especially, it is not applicable to cases of continuous states or actions. 
			Different authors have also reported poor performance due to lack of convergence property in multi-agent settings \citep[e.g.,][]{agogino2005quicker, georgila2014single}. 

			In this paper, an efficient adaptation of Watkins' Q-learning method, called fuzzy Q-learning, is applied to facilitate a continuous state space and action space. 
			Fuzzy Q-learning proposed by \cite{glorennec1994fuzzy} is an universal approximator of the Q-function defined in Eq. \ref{eq:Q_Update_Rule} 
			and combines Q-learning and a fuzzy rule-based system. 
			The fuzzy rule-based system in this investigation has the following form, which can be regarded as a zero order Takagi-Sugeno fuzzy system. 
			\begin{align}
				\text{fuzzy rule }i \hspace{-1mm}: \hspace{2mm} \text{IF } s_{t,S} \text{ IS } L_S^i & \text{ AND } s_{t,B}\text{ IS } L_B^i \text{ THEN } \label{eq:fuzzy_rule_based_system} \\ 
				 & Q_t(\pmb{s}_t, \pmb{a}_t) = \sum\limits_{i=1}^N \alpha_i(\pmb{s}_t) \; q_t[i,a_{t,i}] \label{eq:Q_values} \\ 
				& A_t(\pmb{s}_t, \pmb{a}_t) = \sum\limits_{i=1}^N \alpha_i(\pmb{s}_t) \; a_t[i,a_{t,i}] \label{eq:A_values}
			\end{align}
			The parameter $i \in \{ 1,...,N \}$ describes the index of fuzzy rules, $N \in \mathbb{N}$ is the number of fuzzy rules, $\pmb{s}_t = (s_{t,S},s_{t,B}) \in \mathcal{S} \subset \mathbb{R}^2$ represents the state vector, $\pmb{a}_t=(a_{t,1},...,a_{t,N}) \in \{ 1,...,K \}^N$ denotes the index vector of stored actions $a_t[i, a_{t,i}] \in \mathcal{A} \subset \mathbb{R}$, $K \in \mathbb{N}$ is the number of stored actions in each fuzzy rule, $A_t(\pmb{s}_t, \pmb{a}_t) \in \mathbb{R}$ refers to the inferred action, and $q_t[i,a_{t,i}] \in \mathbb{R}$ denotes the stored q-values.
			As with the conventional Q-learning approach, the q-values are stored in a look-up table, but fuzzy Q-learning requires much less memory storage, since the table size depends on the number of fuzzy rules $N$ and on the number of stored actions $K$ in each fuzzy rule. $N$ and $K$ rely on the fuzzy partition of the state space $\mathcal{S}$ and on the discretization of the action space $\mathcal{A}$, respectively. For the sake of simplicity, it is assumed that each fuzzy rule has exactly $K$ possible actions. 
			
			Note, in fuzzy Q-learning, $\mathcal{S}$ and $\mathcal{A}$ are continuous spaces and the inferred action $A_t(\pmb{s}_t, \pmb{a}_t)$ of the supplying division (buying division) corresponds to the investment decision $I_S$ ($I_B$). 
			Further, the fuzzy sets $L_j^i$ are usually characterized by linguistic labels (e.g., ``low'', ``high'') and the function $\alpha_i(\pmb{s}_t)$ denotes the truth value of rule $i$ given the state vector $\pmb{s}_t$. 
			Finally, Q-values $Q_t(\pmb{s}_t, \pmb{a}_t)$ and actions $A_t(\pmb{s}_t, \pmb{a}_t)$ are inferred from Eq. \ref{eq:Q_values} and Eq. \ref{eq:A_values}, respectively. 
	
			In fuzzy Q-learning, the Q-function is no longer a look-up table, since the Q-values are estimated by a linear parameterized approximation of a few stored q-values, where the ``weights" $\alpha_i(\pmb{s}_t)$ are commonly generated by the T-norm product 
			\begin{equation}
				\alpha_i(\pmb{s}_t) = \mu_{L_S^i}(s_{t,S}) \cdot \mu_{L_B^i}(s_{t,B}) \;, \label{eq:truth_value} 
			\end{equation}
			where $\mu_{L_j^i}(s_{t,j}) \in [0,1]$ denotes the membership function of rule $i$ in state $s_{t,j}$, $j \in \{ S,B \}$. 
			The T-norm (or triangular norm, see, e.g., \cite{zimmermann2011fuzzy}) is a type of binary operation that is often used in fuzzy logic to model the AND operator in Eq. \ref{eq:fuzzy_rule_based_system}. 
			Usually, the membership function (or membership grade) is defined on the interval $[0,1]$, which implies that if an object has a membership grade of one (zero) in a fuzzy set, then the object is absolutely (not) in that fuzzy set. 
			In addition, the membership functions considered here are set in such a way that the strong fuzzy partition is fulfilled, i.e., $\sum_{i=1}^N \alpha_i(\pmb{s}_t) =1$ for each $\pmb{s}_t \in \mathcal{S}$. %
			
			Similar to the conventional Q-learning method, the q-values are updated by 
			\begin{equation}
				q_{t+1}[i,a_{t,i}] = q_t[i,a_{t,i}] + \alpha_i(\pmb{s}_t) \; \Delta Q_t(\pmb{s}_t, \pmb{a}_t) \;,	\label{eq:Fuzzy_Q_Update_Rule} 	
			\end{equation}
			where $\Delta Q_t(\pmb{s}_t, \pmb{a}_t)$ is the temporal difference error which is given by 
			\begin{equation}
				\alpha \; \Big( r_t(\pmb{s}_t,A_t(\pmb{s}_t, \pmb{a}_t)) \; + \; \gamma \; \sum\limits_{i=1}^N \alpha_i(\pmb{s}_{t+1}) \max\limits_{k \in \{ 1,...,K\}} q_t[i,k] \; - \; Q_t(\pmb{s}_t, \pmb{a}_t) \Big) \; . \label{eq:Fuzzy_Q_Error} 	
			\end{equation}
			According to Eq. \ref{eq:Fuzzy_Q_Update_Rule}, up to $N$ q-values per time step can be adjusted and, therefore, the Q-function converges more quickly in fuzzy Q-learning.\footnote{ 
			Roughly speaking, the Q-function converges to its optimum, when a sufficient number of time steps has been performed and the stored q-values change only slightly, or if the temporal difference error $\Delta Q_t(\pmb{s}_t, \pmb{a}_t)$ goes to zero, i.e., $\Delta Q_t(\pmb{s}_t, \pmb{a}_t) \to 0$ for $t \to \infty$. 
			For the precise Q-learning conditions which ensure that the Q-function converges (from any initial state) see, e.g., \cite{watkins1992q}.
			According to the parameter settings in Sec. \ref{sec:Parameter_Settings_and_Simulation_Setup}, up to $4$ q-values are updated simultaneously per time step.} 
			Moreover, fuzzy Q-learning is simple, has low computational demands per iteration, and, with relatively few environmental interactions, good results can be achieved. 
			Nevertheless, there are more sophisticated learning techniques in reinforcement learning (for an overview see, e.g., \cite{sutton2018reinforcement}). 

	\subsection{Exploration policy} 
	\label{ss:Exploration_Polic}




		A very key part in reinforcement learning is the trade-off between exploration and exploitation. 
		In order to select ``good'' actions, an agent has to explore its environment and, in particular, has to learn which action provides which reward. 
		As with other reinforcement learning approaches, a balance between exploration and exploitation has to be found. 
		Therefore, an agent is confronted with the trade-off between choosing the currently optimal action (exploitation) and choosing a varied action with the prospect of a higher reward in the future (exploration) \citep{sutton2018reinforcement}. 
		Nevertheless, the agent needs the possibility to explore all states and all actions frequently enough to ensure the convergence of the Q-function to an optimum. 
		In the following, three exploration policies, which are applied in the simulation study, are explained. 

		\subsubsection{Boltzmann exploration policy}
		\label{sssec:Boltzmann_Exploration_Policy}
		
			A commonly used and very sophisticated exploration policy is the Boltzmann exploration policy \citep[][]{cesa2017boltzmann}.\footnote{
			For other exploration policies see, e.g., \cite{thrun1992efficient}, and, by the way, the Boltzmann exploration policy is sometimes referred to as Gibbs or softmax exploration policy \citep{sutton2018reinforcement}.} 
			For each action $a_t[i,k]$, $i \in \{ 1,...,N\}$ and $k \in \{ 1,...,K\}$, the Boltzmann probability mass function is given by 
			\begin{equation}
				P_{t,i,k}(a_t[i,k]) = \frac{exp\big( {q_t[i,k]/\beta(t)} \big)}{\sum_{l=1}^{K} exp\big( {q_t[i,l]/\beta(t)} \big)} \;, \label{eq:Boltzmann_Probability} 
			\end{equation}
			where $\beta(t) \in \mathbb{R}^+$ denotes the experimentation tendency \citep{waltman2008learning}. 
			According to Eq. \ref{eq:Boltzmann_Probability}, it holds $\sum_{k=1}^{K} P_{t,i,k} = 1$ for each fuzzy rule $i$ and time step $t$. 
			The Boltzmann distribution states that actions with higher q-values are more likely to be selected than actions with lower q-values. 
			The degree of randomness for exploration is controlled by the experimentation tendency \citep{powell2012optimal} and, therefore, $\beta(t)$ should change slowly over time. 
			In particular, if $\beta(t)$ tends to infinity, the shape of the Boltzmann distribution converges to the discrete uniform distribution, which means that all actions have the same probability to being chosen (pure exploration). 
			Contrary, if $\beta(t)$ decreases toward zero, only the action with the highest Q-value is selected (pure exploitation). 
			
			For the long-term behavior of self-learning agents, the experimentation tendency should slowly decrease over time in order to reduce exploration \citep{dearden1998bayesian}. 
			Hence, the experimentation tendency is calculated by a simple rational function	
			\begin{equation}
				\beta(t) = \frac{\beta_1}{\beta_2+t} \; , \label{eq:experimentation_tendency}
			\end{equation}
			whereby the experimentation tendency parameters $(\beta_1,\beta_2) \in \mathbb{R}^2$ should be set in such a way that the learning agents have a long time to explore for good policies, but also time to exploit them.\footnote{
			Rational functions, like Eq. \ref{eq:experimentation_tendency}, have the advantage over linear functions that the experimentation tendency decreases faster at the beginning of a simulation run, but it never reaches zero. 
			Technical note: the experimentation tendency $\beta(t)$ should not become too small during a simulation run, otherwise the expression $exp(q_t[i,k]/\beta(t))$ will be too large and this could endanger the numerical stability of the simulation.} 
			
			
			Note, after all probabilities have been calculated according to Eq. \ref{eq:Boltzmann_Probability}, the index of each stored action is drawn from
			\begin{equation}
				a_{t,i} \thicksim \bigg( {1 \atop P_{t,i,1}},...,{K \atop P_{t,i,K}} \bigg) 
			\end{equation}
			for each fuzzy rule $i$.

		\subsubsection{$\epsilon$-greedy exploration policy}
		\label{sssec:Epsilon_Greedy_Exploration_Policy}
		
			Another common and fairly easy exploration policy is the so-called $\epsilon$-greedy exploration policy, which means that with probability $1-\epsilon(t)$ the action with the highest q-value is chosen, while with probability $\epsilon(t) \in [0,1]$ a random action is selected \citep[][]{sutton2018reinforcement}. 
			For each fuzzy rule $i \in \{ 1,...,N \}$, the index of the stored action is calculated by the following rule: 
			\begin{equation}
				a_{t,i}= \begin{cases}
				\vspace{3mm} \underset{k \in \{ 1,...,K \}}{\mathrm{arg\,max}} q_t[i,k] & \text{with probability $1-\epsilon(t)$} \\
				\thicksim Unif \Big( \{ 1 ,...,K \} \Big) & \text{with probability $\epsilon(t)$}
				\end{cases}
			\end{equation}		
			In the simulation study, $\epsilon(t)$ is a decreasing linear function of the time
			\begin{equation}
				\epsilon(t) = \begin{cases}
				\vspace{3mm}  \epsilon_1 - \epsilon_2 \; t & \text{for $t \leq T_L$} \\
				0 & \text{for $t > T_L$}
				\end{cases}
			\end{equation}	
			with the two properties that, at the beginning of a simulation run, $\epsilon(1)$ is one, which indicates that the action-selection is total random (pure exploration) and, in the end, $\epsilon(T_L)$ is zero (pure exploitation). 
			The parameter $T_L$ is discussed in Sec. \ref{sec:Parameter_Settings_and_Simulation_Setup}. 

		\subsubsection{Upper confidence bound exploration policy}
		\label{sssec:Upper_Confidence_Bound_Exploration_Policy}

			For the third and last exploration policy, the so-called upper confidence bound exploration policy (often abbreviated as the UCB policy) is adapted, which was also proposed for multi-armed bandit problems \citep{auer2002finite}. 
			The idea behind this exploration policy is that not only the current q-values are used for the action-selection, but also the uncertainty that comes along with that selection. 
			One way of doing this is to select an action according to the following rule \citep[][]{sutton2018reinforcement}:
			\begin{equation}
				a_{t,i}= \begin{cases}
				\vspace{3mm} \underset{k \in \{ 1,...,K \}}{\mathrm{arg\,max}} q_t[i,k] + c_1 \sqrt{\frac{ln(t)}{N_t[i,k]}} & \text{if $\forall N_t[i,k] > 0$} \\ 
				\thicksim Unif \Big( \{ k \in \{ 1,...,K \} \; | \; N_t[i,k]=0 \} \Big) & \text{else} 
				\end{cases}
			\end{equation}	
			The parameter $c_1 \in \mathbb{R}$ describes the degree of exploration and $N_t[i,k] \in \mathbb{N}$ is the number of times that the index of the stored action $k$ in fuzzy rule $i$ has been chosen prior to time $t$. 
			At the beginning of a simulation run, all $N_t[i,k]$ are initialized with zero and, each time $a_{t,i}$ is determined, $N_t[i,a_{t,i}]$ is incremented by one, and, hence, the ``uncertainty term'' is reduced. 
			
			Note that actions that have never been chosen are given preference over actions that have already been chosen and the use of the natural logarithm implies that the uncertainty increases slightly over time. 
			If $c_1$ is set too high, then the distribution of the UCB policy converges to the discrete uniform distribution. 
			If $c_1$ is set to zero, then the highest q-value determines which action is chosen. 

\section{Parameter settings and simulation setup}
\label{sec:Parameter_Settings_and_Simulation_Setup}

\begin{figure}[h!]
		\centering
		\includegraphics[width=1\textwidth]{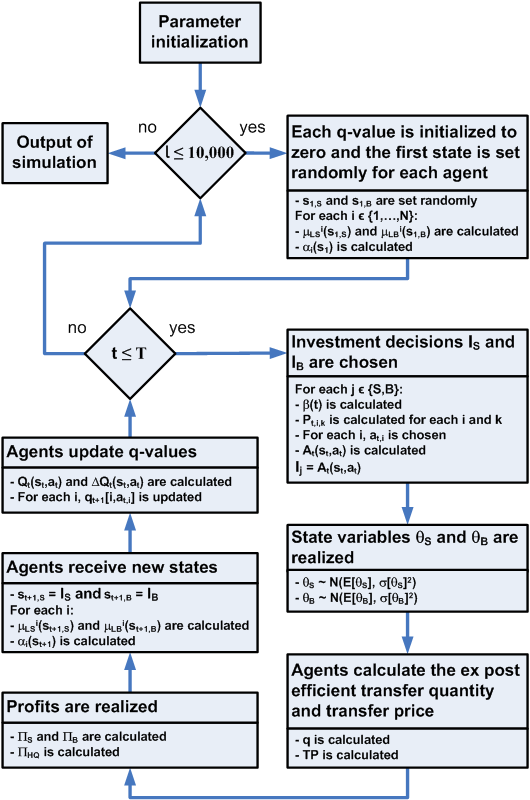}
		\caption{Flow diagram of the agent-based simulation using the Boltzmann exploration policy.} 
		\label{fig:Procedure}
	\end{figure}

	The agent-based simulation is conducted in four main steps: First, the scenario in which both divisions have the same marginal cost parameter is studied, i.e., $\lambda_S=\lambda_B$ and, according to Eq. \ref{eq:Gamma_sb}, this yields to a surplus sharing parameter $\Gamma$ of $0.5$, which means that each division receives the same fraction of the contribution margin achieved. 
	In the next step, the case where the divisional investment costs are not symmetrical is examined, which means that the contribution margin is not divided symmetrically either. 
	A third analysis studies whether the divisions make higher investment decisions, if the surplus sharing parameter $\Gamma$ has different values than the theory of subgame perfection equilibrium recommends. 
	In the first three investigations, the standard deviations of the state variables is set to zero, which implies that there are no stochastic fluctuations that affect the decision-making process, and, in addition, it is supposed that the divisions explore their environment using the Boltzmann exploration policy. 
	In the fourth and final step, a sensitivity analysis with regard to the standard deviations of the state variables and the exploration policies is carried out in order to check whether the observed results also occur in stochastic environments with different exploration settings.\footnote{ 
	A simulation study should include a systematic  sensitivity analysis to verify the stability of the simulation outcomes \citep[e.g.][]{davis2007developing, guerrero2011using}. 
	}
	For a better understanding, the parameter settings are successively explained in more detail and  the procedure of the agent-based simulation is illustrated (the exogenous parameters are summarized in Table \ref{tab:Parameter_Setting}, while the procedure of the agent-based simulation is depicted in Fig. \ref{fig:Procedure}). 

	\begin{table}[h!]
	\linespread{1.1}\selectfont
	\centering
	\caption{Parameter settings}\label{tab:Parameter_Setting}
	\begin{tabular}{|l|l|l|}
		\hline	
		\textbf{Fixed exogenous parameters} \hspace{3mm} & \textbf{Values} \\
		\hline
		Number of simulation runs & $10,\hspace{-0.6mm}000$ \\
		Number of time steps per simulation run & $T=1,\hspace{-0.6mm}100$ \\ 
		Time steps to learn the Q-function & $T_L=1,\hspace{-0.6mm}000$ \\
		Time steps to evaluate the outcome &  $T_E=100$ \\
		Slope of inverse demand function & $b=12$ \\
		Expected values of the state variables & $E[\theta_S] = 60$ and $E[\theta_B] = 100$ \\
		Action space & $\mathcal{A}=\{ 0,5,...,50 \}$ \\
		State space & $\mathcal{S} \hspace{0.5mm} =\{ 0,12.5,...,50 \}\hspace{-1mm} \times \hspace{-1mm} \{ 0,12.5,...,50 \}$ \\
		Number of fuzzy rules & $N=25$ \\
		Number of stored actions in each fuzzy rule & $K=11$ \\
		Learning rate & $\alpha =0.5$ \\
		\hline
		\hline
		\textbf{Scenario-based exogenous parameters} & \textbf{Values} \\
		\hline
		Marginal cost parameter & $\lambda_S \in \{ 1/2, \hspace{0.3mm} 7/12, \hspace{0.3mm} 2/3, \hspace{0.3mm} 3/4, \hspace{0.3mm} 5/6 \}$ \\
		 & \hspace{3.2mm} $\approx\{ 0.5, 0.58, 0.67, 0.75, 0.83 \}$ \\
		 & $\lambda_B = 1-\lambda_S$ \\
		Supplying division's surplus sharing parameter & $\Gamma \in \{ 0.1, 0.2,...,0.9 \}$ \\
		Discount factor & \hspace{-0.4mm} $\gamma \in \{ 0.0,0.1,...,0.9 \}$ \\	
		Standard deviations of the state variables & $\sigma[\theta_S]=\sigma[\theta_B] \in \{ 0, 5,10 \} $ \\
		Exploration policies: & \\ 
		\hspace{3mm} Boltzmann exploration policy & $\beta_1=12487.5$ and $\beta_2=248.75$ \\
		\hspace{3mm} $\epsilon$-greedy exploration policy & $\epsilon_1=1\hspace{-0.6mm}+\hspace{-0.6mm}1/999$ and $\epsilon_2=-1/999$ \\
		\hspace{3mm} Upper confidence bound exploration policy & $c_1=30$ \\
		\hline
	\end{tabular}
	\linespread{1}\selectfont
	\end{table}
	
	This simulation study focuses on settings in which the learning behavior of agents is modeled by fuzzy Q-learning. 
	It is assumed that each division has an individual constant marginal cost parameter $\lambda_j$, $j \in \{ S,B\}$. 
	Since there are infinitely many combinations that can be examined, the simulation experiments only focus on the following convex combination, namely, $\lambda_S+\lambda_B=1$ with $\lambda_j \geq 0$. 
	In order to analyze scenarios with surplus sharing parameters $\Gamma$ ranging from $0.1$ to $0.9$ in steps of $0.1$ (and in order that the first-best and the second-best case have integer solutions for $\Gamma=0.5$), the slope of the inverse demand function $b$ is set to $12$ as well as the expected values of the state variables $\theta_S$ and $\theta_B$ is set to $60$ and $100$, respectively. 
	The corresponding $\lambda_j$ values are given in Table \ref{tab:Parameter_Setting}, whereas the solutions of subgame perfect equilibrium are listed in Table \ref{tab:Firstbest_Secondbest_Solutions}. 
	By the way, scenarios where $\Gamma$ is one (zero) lead, according to Eq. \ref{eq:optimal_I_j} and Eq. \ref{eq:second_best_I_j}, to the result that the supplying division (buying division) chooses first-best investments, but the other division has no incentive to invest as any investment cause divisional investment costs and, at the same time, it does not benefit from the resulting profit. 
	Hence, such scenarios are less interesting to analyze 
	and, due to the fact that both divisions have the same quadratic cost structure (cf. Eq. \ref{eq:divisional_investment_costs}) and that $\lambda_B=1-\lambda_S$, it is sufficient to study scenarios for $\lambda_S$ ranging from $0.5$ to $0.83$. 
	
	\begin{table}[h!]
	\linespread{1.1}\selectfont
	\centering
	\caption{First-best and second-best solutions resulting from the concept of subgame perfect equilibrium for $b=12$, $E[\theta_S]=60$, and $E[\theta_B]=100$.}
	\label{tab:Firstbest_Secondbest_Solutions}
	\begin{tabular}{|ccc|cccccc|}
		\hline	
		\multicolumn{9}{|l|}{\textbf{First-best solutions}} \\
		\hline
		$\Gamma$ & $\lambda_S$ & $\lambda_B$ & $I_S^*$ & $I_B^*$ & $q^*$ & $\Pi_S^*$ & $\Pi_B^*$ & $\Pi_{HQ}^*$ \\
		\hline
		$0.1$ & $0.83$ & $0.17$ & $10$ & $50$ & $8.33$ & $0$ & $166.67$ & $166.67$ \\
		$0.2$ & $0.75$ & $0.25$ & $8$ & $24$ & $6$ & $19.2$ & $100.8$ & $120$ \\
		$0.3$ & $0.67$ & $0.33$ & $8$ & $16$ & $5.33$ & $29.87$ & $76.8$ & $106.67$ \\
		$0.4$ & $0.58$ & $0.42$ & $8.70$ & $12.17$ & $5.07$ & $39.70$ & $61.75$ & $101.45$ \\
		$0.5$ & $0.5$ & $0.5$ & $10$ & $10$ & $5$ & $50$ & $50$ & $100$ \\
		\hline
		$0.6$ & $0.42$ & $0.58$ & $12.17$ & $8.70$ & $5.07$ & $61.75$ & $39.70$ & $101.45$ \\
		$0.7$ & $0.33$ & $0.67$ & $16$ & $8$ & $5.33$ & $76.8$ & $29.87$ & $106.67$ \\
		$0.8$ & $0.25$ & $0.75$ & $24$ & $8$ & $6$ & $100.8$ & $19.2$ & $120$ \\
		$0.9$ & $0.17$ & $0.83$ & $50$ & $10$ & $8.33$ & $166.67$ & $0$ & $166.67$ \\
		\hline
		\hline	
		\multicolumn{9}{|l|}{\textbf{Second-best solutions}} \\
		\hline
		$\Gamma$ & $\lambda_S$ & $\lambda_B$ & $I_S^{sb}$ & $I_B^{sb}$ & $q^{sb}$ & $\Pi_S^{sb}$ & $\Pi_B^{sb}$ & $\Pi_{HQ}^{sb}$ \\
		\hline
		$0.1$ & $0.83$ & $0.17$ & $0.74$ & $33.33$ & $6.17$ & $22.63$ & $113.17$ & $135.80$ \\
		$0.2$ & $0.75$ & $0.25$ & $1.25$ & $15$ & $4.69$ & $25.78$ & $77.34$ & $103.13$ \\
		$0.3$ & $0.67$ & $0.33$ & $1.90$ & $8.89$ & $4.23$ & $31.04$ & $62.08$ & $93.12$ \\
		$0.4$ & $0.58$ & $0.42$ & $2.78$ & $5.83$ & $4.05$ & $37.13$ & $51.99$ & $89.12$ \\
		$0.5$ & $0.5$ & $0.5$ & $4$ & $4$ & $4$ & $44$ & $44$ & $88$ \\
		\hline
		$0.6$ & $0.42$ & $0.58$ & $5.83$ & $2.78$ & $4.05$ & $51.99$ & $37.13$ & $89.12$ \\
		$0.7$ & $0.33$ & $0.67$ & $8.89$ & $1.90$ & $4.23$ & $62.08$ & $31.04$ & $93.12$ \\
		$0.8$ & $0.25$ & $0.75$ & $15$ & $1.25$ & $4.69$ & $77.34$ & $25.78$ & $103.13$ \\
		$0.9$ & $0.17$ & $0.83$ & $33.33$ & $0.74$ & $6.17$ & $113.17$ & $22.63$ & $135.80$ \\
		\hline
	\end{tabular}
	\linespread{1}\selectfont
	\end{table}	
	
	Next, the choice of the action space $\mathcal{A}$ is discussed. 
	If the number of possible investment decisions is too small, the learning behavior of both divisions cannot be thoroughly investigated, and, on the other hand, if the size of the action space is chosen too large, then learning is slowed. 
	In order to ensure that the first-best investment decisions are in the action space in all examined scenarios, the action space must have at least a minimum size of $50$. 
	Since it can be guessed that a human decision-maker can only differentiate between a limited number of actions, the stored actions $a_t[i,k] \in \mathcal{A}$ vary between $0$ and $50$ in steps of $5$. 
	Consequently, there are $11$ stored actions for each fuzzy rule $i \in \{ 1,...,N \}$. 
	Recall that, according to Eq. \ref{eq:A_values}, regardless of whether the stored actions have integer values or not, the inferred action $A_t$ (which corresponds to the agent's investment decision) is a real number. 

\begin{figure}[h!]
		\centering
		\includegraphics[width=0.85\textwidth]{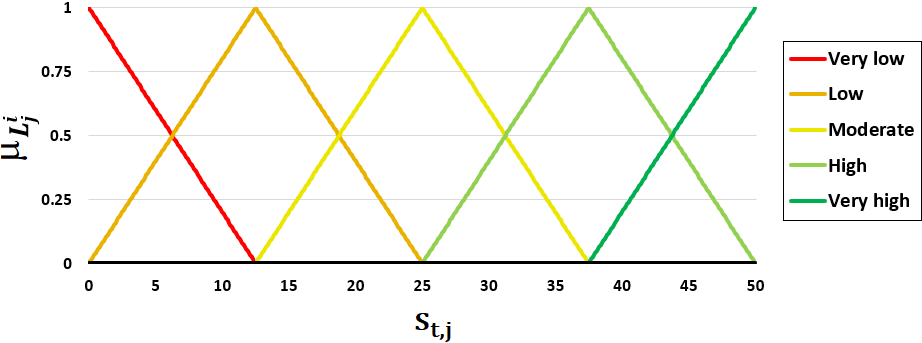}
		\caption{The fuzzy sets considered here consist of $5$ triangular membership functions in each state space dimension $j \in \{S,B\}$, which can be characterized by linguistic labels ranging from ``Very low'' to ``Very high''.}
		\label{fig:Triangular_Membership_Functions}
	\end{figure}

	Also for the state space $\mathcal{S}$, it is supposed that the number of fuzzy rules should be small. 
	In order to deal with $5$ membership functions in each state space dimension, the values of the state space range from $0$ to $50$ by a step size of $12.5$; thus, there are $25$ fuzzy rules in total. 
	According to \cites{zimmermann2011fuzzy} suggestions, triangular membership functions are applied (cf. Fig. \ref{fig:Triangular_Membership_Functions}) because they are simple, have low computing power per iteration, and are easy to interpret. 
	For example, if the state of the supplying division $s_{t,S}$ is $10$, then the membership grade $\mu_{L_S^1}$ of fuzzy rule $1$ is $0.2$, whereas the membership grade $\mu_{L_S^2}$ of fuzzy rule $2$ is $0.8$, and all other membership grades $\mu_{L_S^3}$, $\mu_{L_S^4}$, and $\mu_{L_S^5}$ are zero, or in other words, the supplying division is in the ``very low investment range'' with a fraction of $20\%$ and in the ``low investment range'' with a share of $80\%$. 
	Incidentally, increasing the number of fuzzy rules in the parameter settings just means that the agents need more time to learn their environment; their decision-making behavior can be exhibited quite well with only $5$ membership functions. 
	Note, according to the triangular membership functions in Fig. \ref{fig:Triangular_Membership_Functions}, up to $4$ q-values can be updated simultaneously per time step ($2$ raised to the power of $dim(\mathcal{S})$). 

	Furthermore, the research is limited to just one learning rate, namely, $\alpha =0.5$, because it can be assumed that, for long-term decisions, old stored information is just as important as new received information.  
	Regardless of the selection of the learning rate, the Q-function converges (sometimes faster and sometimes slower) to an optimum with the parameter settings giving in Table \ref{tab:Parameter_Setting}. 
	Note, if the learning rate is set too high, the learned information will be regularly overwritten, which is usually not recommended as this requires more time steps for convergence and, in particular, stochastic fluctuations during learning make convergence difficult.
	On the other hand, if the learning rate is chosen too small, many more time steps are required for convergence. 
	
	In the agent-based simulation, the number of time steps to learn the Q-function $T_L \in \mathbb{N}$ is mainly based on the cardinalities of $\mathcal{A}$ and $\mathcal{S}$; other factors like $\alpha$, the degree of exploration, and the extent of stochastic fluctuations also influence the learning time, but to take them into account explicitly would be too time-consuming. 
	Since each agent has $275$ ($=|\mathcal{A}| \cdot |\mathcal{S}|^2=11 \cdot 5^2$) different q-values and the maximum number of q-values that are updated per time step is $4$, 
	 the number of time steps to learn the Q-function is set to $1,\hspace{-0.6mm}000$ so that a sufficient number of state-action pairs are visited. 
	The pre-generated simulations also suggest that a learning time of $1,\hspace{-0.6mm}000$ is sufficient to guarantee convergence in all examined scenarios. 
	After determining the number of time steps required for learning the Q-function, further $100$ time steps are simulated to evaluate the agents' outputs, hence $T_E=100$ and, consequently, the number of time steps per simulation run $T$ is set to $1,\hspace{-0.6mm}100$.\footnote{ 
	Please be aware that even if the Q-function converges, the stochastic fluctuations of the environment affect the divisions' profits. In order to take the volatility of the markets into account, an observation period of $100$ time steps is chosen to evaluate the simulation outcomes.} 
	
	It should be emphasized once again that this study aims to analyze negotiated transfer pricing with cognitively bounded agents. 
	This means that not only the number of actions or the number of fuzzy rules is limited, but also the possibility of observing all state-action pairs infinitely often. 
	Therefore, the simulation ends after $1,\hspace{-0.6mm}100$ time steps, even if not all state-action pairs have been visited. 
	
	Besides the surplus sharing parameter $\Gamma$, also the discount factor $\gamma$ varies from $0$ to $0.9$ in steps of $0.1$. 
	The discount factor is set exogenously at the very beginning of a simulation run and represents the agent's foresight. 
	A discount factor of zero indicates that the agent is very myopic, which implies that the agent cannot observe future rewards. 
	Conversely, if $\gamma$ goes to one, the agent becomes more non-myopic, so the agent will place more emphasis on future rewards. 
	Recall, when $\gamma$ is set to one, then the Q-function does not converge. 
	
	Finally, the state variables $\theta_S$ and $\theta_B$ are captured in a normally distributed random variable with mean $60$ and $100$, respectively. 
	To describe the turbulence of the environment in which the agents operate, the standard deviations of the state variables are set to $0$, $5$, and $10$, which can be interpreted as a deterministic environment, an environment with minor stochastic fluctuations, and an environment with considerable volatility on the markets. 

	In addition to varying the state variables, the sensitivity of the results is checked in regard to three different exploration policies. 
	Inspired by the work of \cite{tijsma2016comparing}, this paper focuses on the Boltzmann, $\epsilon$-greedy, and upper confidence bound exploration policy. 
	For the introduced Boltzmann exploration policy in Sec. \ref{sssec:Boltzmann_Exploration_Policy}, one has to define the exploration tendency parameters $\beta_1$ and $\beta_2$ in Eq. \ref{eq:experimentation_tendency}. 
	The pre-generated simulations reveal that an experimentation tendency $\beta(t)$ of around $50$ at the beginning of a simulation run is a good choice because the agents are able to explore many action-state pairs. 
	On the other hand, a value of $10$ at the end of learning is sufficient for the convergence of the Q-function and, from a technical point of view, $\beta(t)$ ranges in a value domain in which no numerical instabilities occur during a simulation run.  
	Solving Eq. \ref{eq:experimentation_tendency} with the two associated boundary conditions ($\beta(1)=50$, and $\beta(T_L)=10$) leads to the Boltzmann exploration tendency parameters given in Table \ref{tab:Parameter_Setting}. 
	Setting the exploration parameters $\epsilon_1$ and $\epsilon_2$ for the $\epsilon$-greedy exploration policy is simple because $\epsilon(1)=1$ and $\epsilon(T_L)=0$. 
	The associated $\epsilon$-greedy exploration parameters are reported in Table \ref{tab:Parameter_Setting}. 
	Finally, the UCB exploration parameter $c_1$ has to specify. 
	According to the pre-generated simulations, a value of about $30$ is a good choice for $c_1$. 
	Nevertheless, balancing the parameters of an exploration policy is one of the most challenging tasks in reinforcement learning \citep{tijsma2016comparing}. 
	
	Since the first state of each fuzzy Q-learning agent is determined by a uniform random number generator and the choice of action is based on stochastic exploration policies, each scenario is carried out $10,\hspace{-0.6mm}000$ times and, due to the coefficient of variation (the ratio of the standard deviation to the mean), $10,\hspace{-0.6mm}000$ simulation runs are sufficient to express the precision and repeatability of the simulation study. 

\section{Results and discussion}
\label{sec:Results_and_Discussion}

	This section analyzes the results of the agent-based simulation in four main steps: (1) the results for symmetric marginal cost parameters is presented, i.e., $\lambda_S=\lambda_B=0.5$ and, consequently, $\Gamma=0.5$. 
	(2) the scenario in which the marginal cost parameters are not symmetrical is examined. 
	For this purpose, the case where $\lambda_S$ is $0.83$ is selected, which implies a $\Gamma$ of $0.1$. 
	(3) different surplus sharing parameters $\Gamma$ are tested to see what effect they have on the divisions' investment decisions and the headquarters' profit 
	and (4) an extensive sensitivity analysis is performed to verify whether the results are robust to changes in volatility of the markets and exploration policy. 
	An overview of the investigated scenarios is given in Table \ref{tab:Scenario_overview}. 
	\begin{table}[h!]
	\linespread{1.1}\selectfont
	\centering
	\caption{Scenario overview.}
	\label{tab:Scenario_overview}
	\begin{tabular}{|c|ccccc|}
		\hline
		 & Seller's mar- & Seller's sur- & Discount & Standard & \\
		Scenario & ginal cost & plus sharing & factor & deviation & Exploration policy \\ 
		 & $\lambda_S$ & $\Gamma$ & $\gamma$ & $\sigma[\theta_S]$ & \\
		\hline
		$1$ & $0.5$ & $0.5$ & $(0,0.9)$ & $0$ & Boltzmann \\ 
		$2$ & $0.83$ & $0.1$ & $(0,0.9)$ & $0$ & Boltzmann \\ 
		$3$ & $(0.5,0.83)$ & $(0.1,...,0.9)$ & $(0,...,0.9)$ & $0$ & Boltzmann \\ 
		$4$ & $(0.5,...,0.83)$ & $(0.1,...,0.9)$ & $(0,...,0.9)$ & $(0,5,10)$ & (BM, Greedy, UCB) \\ 
		\hline
	\end{tabular}
	\linespread{1}\selectfont
	\end{table}

	\subsection{Results for symmetric marginal cost parameters}
	
		In a first step, the scenario where each division has the same marginal cost parameter is investigated. 
		According to Eq. \ref{eq:Gamma_sb}, the headquarters assigns equal bargaining power to both divisions. 
		The solutions resulting from the concept of subgame perfection equilibrium are reported in Table \ref{tab:Firstbest_Secondbest_Solutions}. 
		Recall that the first-best solution can be considered as the best-case scenario in which both divisions make optimal decisions. 
		In addition, the standard deviations of the state variables are set to zero so that the agents' decision-making process is not influenced by stochastic environmental fluctuations and, further, the Boltzmann exploration policy is applied for the action-selection of investments. 
	
		In order to get a better understanding of the decision-making behavior of fuzzy Q-learning agents, the simulation study starts with the case of divisions' investment decisions with different levels of foresight. 
		Recall, the agent's foresight is represented by the discount factor $\gamma$ in the fuzzy Q-learning approach. 
		A discount factor of zero indicates that the agent is absolutely myopic and, therefore, does not take any future reward into account. 
		On the other hand, a discount factor close to one implies that the agent strives for a long-term high reward. 

		\begin{figure}[h!]
			\centering
			\includegraphics[width=1\textwidth]{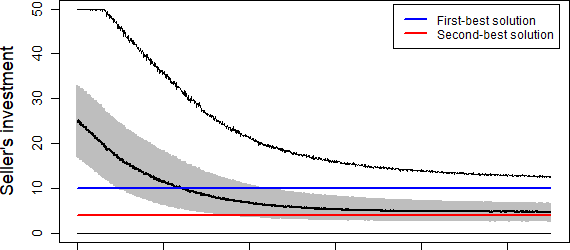} \\
			\vspace{3mm}
			\includegraphics[width=1\textwidth]{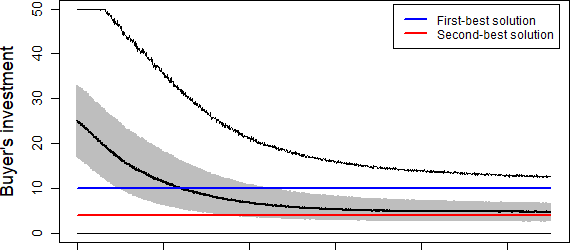} \\
			\vspace{2.5mm}
			\includegraphics[width=1\textwidth]{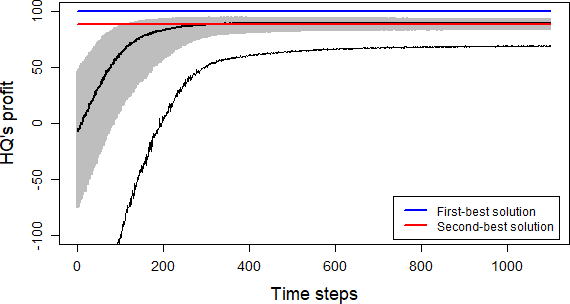}
			\caption{Modified boxplots of seller's investment $I_S$, buyer's investment $I_B$, and headquarters' profit $\Pi_{HQ}$ for myopic fuzzy Q-learning agents in the scenario $1$, i.e., $\lambda_S=\lambda_B=0.5$, $\Gamma=0.5$, $\gamma=0$, $\sigma[\theta_S]=\sigma[\theta_B]=0$, and exploration policy is Boltzmann.}
			\label{fig:Plot_01}
		\end{figure}

		\begin{figure}[h!]
			\centering
			\includegraphics[width=1\textwidth]{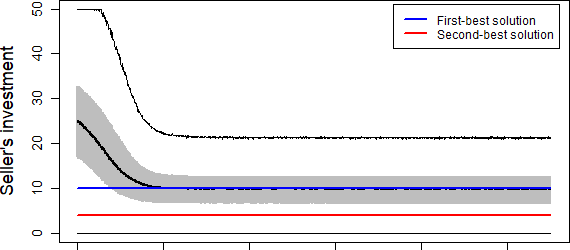} \\
			\vspace{3mm}
			\includegraphics[width=1\textwidth]{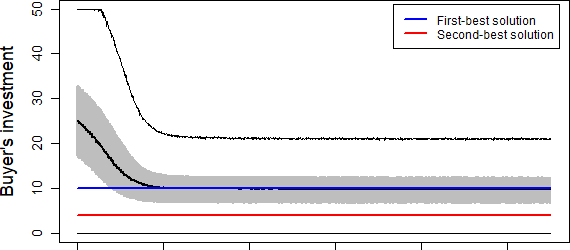} \\
			\vspace{2.5mm}
			\includegraphics[width=1\textwidth]{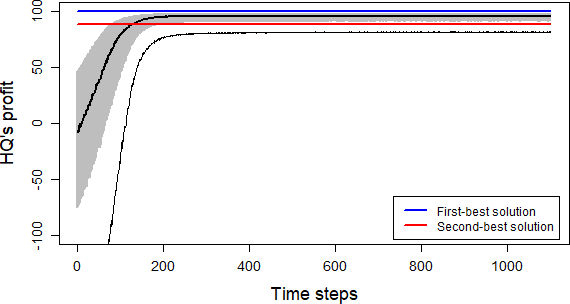}
			\caption{Modified boxplots of seller's investment $I_S$, buyer's investment $I_B$, and headquarters' profit $\Pi_{HQ}$ for non-myopic fuzzy Q-learning agents in the scenario $1$, i.e., $\lambda_S=\lambda_B=0.5$, $\Gamma=0.5$, $\gamma=0.9$, $\sigma[\theta_S]=\sigma[\theta_B]=0$, and exploration policy is Boltzmann.}
			\label{fig:Plot_02}
		\end{figure}

		Fig. \ref{fig:Plot_01} and \ref{fig:Plot_02} depict the divisions' investment decisions and the related headquarters' profit for myopic and non-myopic fuzzy Q-learning agents, respectively. 
		While myopic agents invest only about as much as in the classic hold-up problem, non-myopic agents invest optimally. 
		Since there is no stochastic influence in this scenario setting, the headquarters' profit depends solely on the divisions' investment decisions. 
		A closer analysis reveals that the investment decisions of myopic (non-myopic) agents, averaged over the last $100$ selected investments of $10,\hspace{-0.6mm}000$ simulation runs, are normally distributed with mean $5.04$ ($9.59$) and standard deviation $0.43$ ($3.17$); compared to the second-best solution, fully individual rational utility maximizers only invest an amount of $4$. 
		Further for myopic (non-myopic) agents, the headquarters' profit is almost normally distributed with mean $87.9$ ($93.48$) and standard deviation $1.31$ ($5.14$) and, especially, has a slightly negative skewness of about $-0.29$ ($-1.41$). 
		The findings suggest that, not only from the view of the headquarters, it is better when the divisions act with a high degree of foresight, it is also better for both divisions to seek for high future rewards. 
		Ultimately, it is natural for competing divisions to take expected future rewards into account to maximize their own profits. 
		
		Be aware that the thick black line in the modified boxplots represents the median of $10,\hspace{-0.6mm}000$ observations per time step and, due to the fact that the mean value is not robust against outliers, the mean value of the headquarters' profit is below the median value in case of a negative skewness. 
		Additionally, note that, in the modified boxplots used here, the distance between the 25th and 75th percentile is visualized in gray, the thin black lines denote the whiskers, and, for improved readability, outliers beyond the wiskers are not depicted. 

	\subsection{Results for non-symmetric marginal cost parameters}
	
\begin{figure}[h!]
			\centering
			\includegraphics[width=1\textwidth]{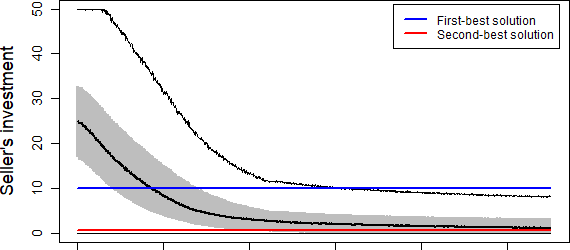} \\
			\vspace{3mm}
			\includegraphics[width=1\textwidth]{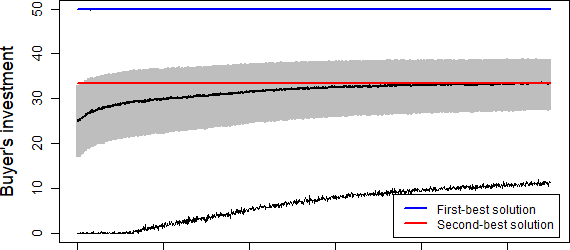} \\
			\vspace{3mm}
			\includegraphics[width=1\textwidth]{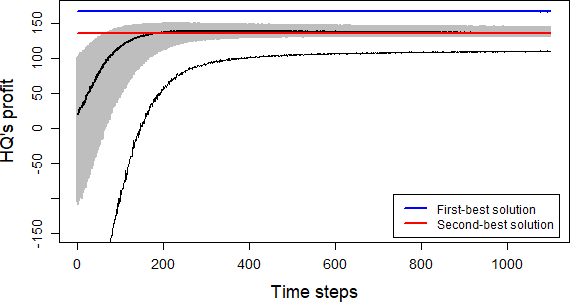}
			\caption{Modified boxplots of seller's investment $I_S$, buyer's investment $I_B$, and headquarters' profit $\Pi_{HQ}$ for myopic fuzzy Q-learning agents in the scenario $2$, i.e., $\lambda_S=0.83$, $\lambda_B=0.17$, $\Gamma=0.1$, $\gamma=0$, $\sigma[\theta_S]=\sigma[\theta_B]=0$, and exploration policy is Boltzmann.}
			\label{fig:Plot_03}
		\end{figure}

\begin{figure}[h!]
			\centering
			\includegraphics[width=1\textwidth]{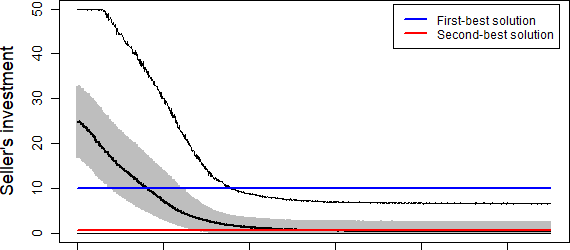} \\
			\vspace{3mm}
			\includegraphics[width=1\textwidth]{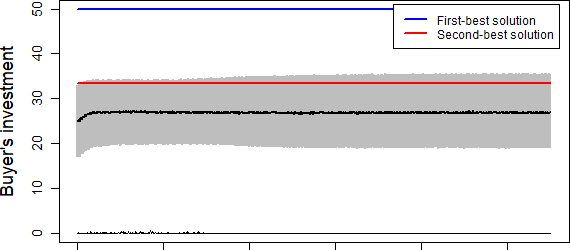} \\
			\vspace{3mm}
			\includegraphics[width=1\textwidth]{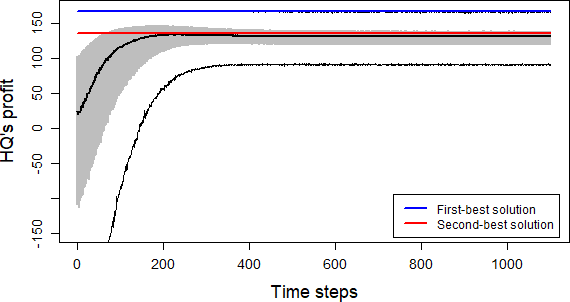}
			\caption{Modified boxplots of seller's investment $I_S$, buyer's investment $I_B$, and headquarters' profit $\Pi_{HQ}$ for non-myopic fuzzy Q-learning agents in the scenario $2$, i.e., $\lambda_S=0.83$, $\lambda_B=0.17$, $\Gamma=0.1$, $\gamma=0.9$, $\sigma[\theta_S]=\sigma[\theta_B]=0$, and exploration policy is Boltzmann.}
			\label{fig:Plot_04}
		\end{figure}

		In the next step, the case where the divisions' marginal cost parameters are not symmetrical is analyzed. 
		According to the selected convex combination, $\lambda_S+\lambda_B=1$, there are four possible scenarios with non-symmetric marginal cost parameters, of which the scenario with $\lambda_S=0.83$ is examined. 
		In the case that the seller's marginal cost parameter is greater than that of the buyer, the headquarters assigns the seller a lower bargaining power (cf. Eq. \ref{eq:Gamma_sb}). 
		For myopic agents, this recommended asymmetrical distribution of bargaining power leads to investments in the area of the second-best solution (cf. Fig. \ref{fig:Plot_03}). 
		In the case of non-myopic agents (cf. Fig. \ref{fig:Plot_04}), the supplying division also invest in the area of the second-best solution, but the buying division makes investment decisions that are even significantly below the second-best solution. 
		In detail, the simulation results show that, for $\gamma=0$, the seller's (buyer's) investments are normally distributed with mean $1.93$ ($32.85$) and standard deviation $0.52$ ($3.47$). 
		For $\gamma=0.9$, the seller's investments are approximately gamma distributed with shape parameter $1.10$ and scale parameter $1.29$, whereas the buyer's investments are approximately normally distributed with mean $26.94$ and standard deviation $8.54$. 
		Additionally, the headquarters' profit resulting from myopic divisions is normally distributed with mean $136.87$ and standard deviation $4.21$, while, for non-myopic divisions, $\Pi_{HQ}$ is approximately normally distributed with mean $127.51$ and standard deviation $13.43$. 
		According to the simulation outputs, the headquarters achieves higher profits with myopic agents than with non-myopic agents. 

		The results for non-symmetrical marginal cost parameters indicate that divisions that have a high discount factor still invest much less than divisions that exhibit a low discount factor. 
		However, the resulting question is what causes this negative effect, as foresight is usually a positive learning characteristic. 
		To gain some insights on the effects of $\gamma$ given $\Gamma$, additional scenarios are considered in which the discount factor $\gamma$ is fixed and the surplus sharing parameter $\Gamma$ is varied in small steps. 

\newpage
	\subsection{Effects of different surplus sharing parameters}

\begin{figure}[h!]
		\centering
		\includegraphics[width=1\textwidth]{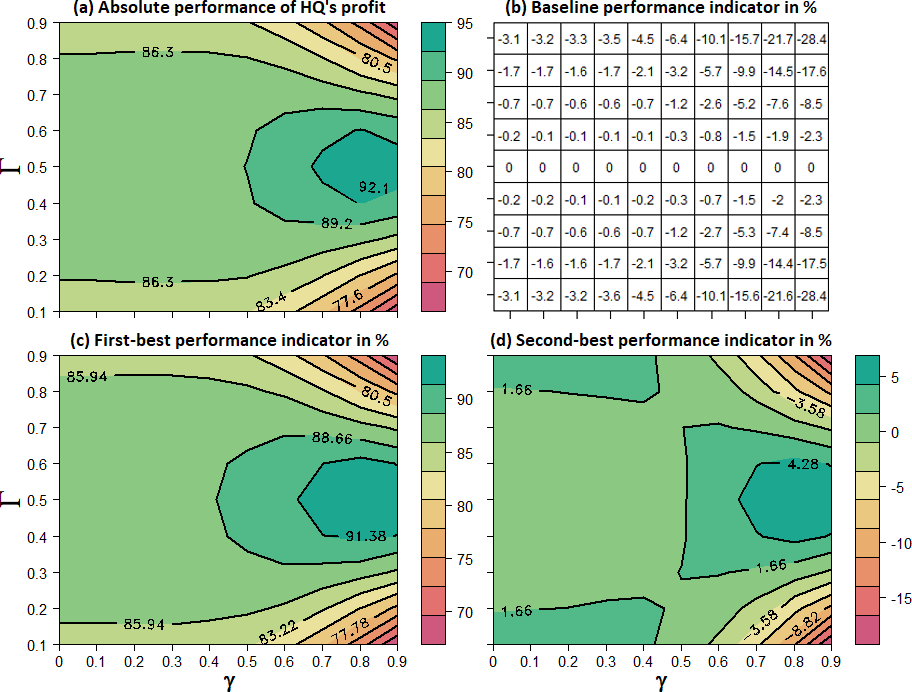}
		\caption{Results for symmetric marginal cost parameters ($\lambda_S=0.5$ and $\lambda_B=0.5$) in the scenario $3$: (a) absolute performance of the headquarters' profit resulting from the simulation, (b) relative performance change and statistical hypothesis testing of $\Pi_{HQ}$ compared to the simulation output from the baseline scenario with $\Gamma^{sb}=0.5$, (c) relative performance of $\Pi_{HQ}$ compared to $\Pi_{HQ}^*$, and (d) relative performance change of $\Pi_{HQ}$ compared to $\Pi_{HQ}^{sb}$.} 
		\label{fig:Plot_05}
	\end{figure}

\begin{figure}[h!]
		\centering
		\includegraphics[width=1\textwidth]{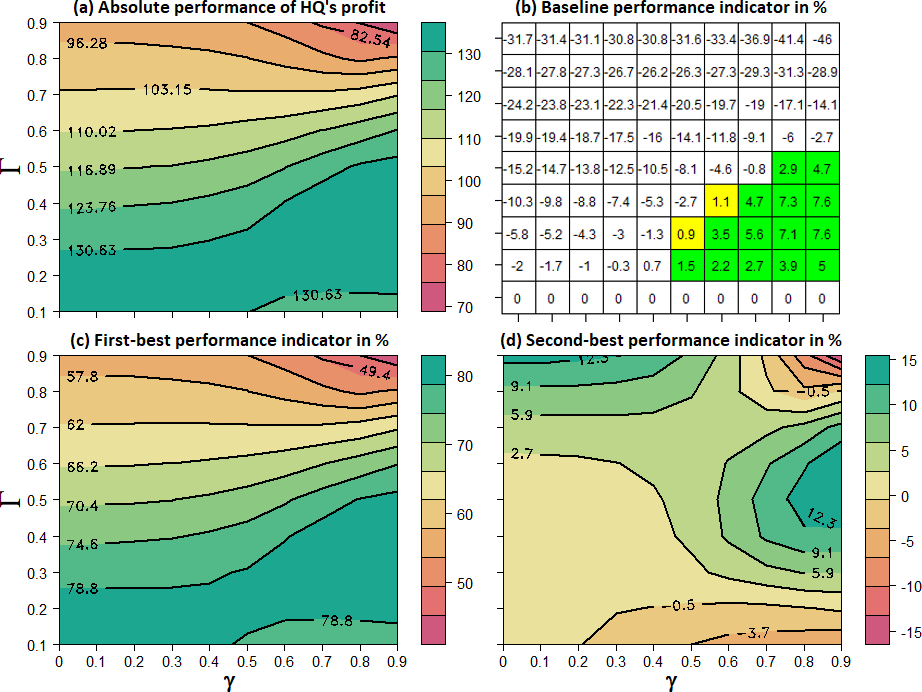}
		\caption{Results for non-symmetric marginal cost parameters ($\lambda_S=0.83$ and $\lambda_B=0.17$) in the scenario $3$: (a) absolute performance of the headquarters' profit resulting from the simulation, (b) relative performance change and statistical hypothesis testing of $\Pi_{HQ}$ compared to the simulation output from the baseline scenario with $\Gamma^{sb}=0.1$, (c) relative performance of $\Pi_{HQ}$ compared to $\Pi_{HQ}^*$, and (d) relative performance change of $\Pi_{HQ}$ compared to $\Pi_{HQ}^{sb}$.}
		\label{fig:Plot_06}
	\end{figure}

		In order to figure out what effects do other $\Gamma$ constellations have on the hold-up problem, further simulations are conducted by varying the surplus sharing parameter in small steps. 
		In addition, the value range for the discount factor is expanded to cover a broader spectrum of the agent's foresight. 
		The simulation outputs are summarized in Fig. \ref{fig:Plot_05} and Fig. \ref{fig:Plot_06} for $\lambda_S=0.5$ and $\lambda_S=0.83$, respectively. 
		In Fig. \ref{fig:Plot_05} and \ref{fig:Plot_06}, there are four subplots to show the impact of $\gamma$ given $\Gamma$ on the simulation outcomes. 
		In the following, the meaning of the subplots is explained. 
	
		In each subplot, the discount factor $\gamma$ is varied on the x-axis from $0$ to $0.9$ in steps of $0.1$, while the surplus sharing parameter $\Gamma$ is varied on the y-axis from $0.1$ to $0.9$ in steps of $0.1$. 
		The top left subplot displays contours representing the headquarters' profits resulting from the simulation. 
		The bottom left subplot depicts the ``first-best performance indicator'' for the fuzzy Q-learning agents, which is defined by $\Pi_{HQ} / \Pi_{HQ}^*$, i.e., the headquarters' profit obtained from the simulation is normalized by the highest feasible profit that can be achieved. 
		The first-best performance indicator can serve as a relative indicator for the profit-effectiveness of the decision-making behavior of the fuzzy Q-learning agents; the higher the value, the better the profitability for the headquarters. 
		For example, if the first-best performance indicator reaches one, then divisions make optimal investment decisions and, in this way, the divisions as well as the headquarters achieve the maximum profit. 
	
		Another performance indicator is visualized in the bottom right subplot. 
		This performance indicator reflects how much better the fuzzy Q-learning agents perform than fully rational utility maximizers do. 
		This called ``second-best performance indicator'' is formalized by the relative change between $\Pi_{HQ}$ and $\Pi_{HQ}^{sb}$ dividing by $\Pi_{HQ}^{sb}$. 
		Thus, the higher the relative change, the higher the performance of fuzzy Q-learning agents. 
		
		Lastly, and more importantly, the so-called ``baseline performance indicator'', which is displayed in the top right subplot, is based on the relative change between $\Pi_{HQ}^{\Gamma}$ and $\Pi_{HQ}^{baseline}$ dividing by $\Pi_{HQ}^{baseline}$, where $\Pi_{HQ}^{\Gamma}$ denotes the headquarters' profit resulting from scenario where the headquarters uses $\Gamma$, while $\Pi_{HQ}^{baseline}$ describes the headquarters' profit resulting from scenario where the headquarters applies $\Gamma^{sb}$. 
		The scenario with $\Gamma^{sb}$ is called the ``baseline scenario''. 
		So in a baseline scenario, the headquarters sets the bargaining power $\Gamma$ according to Eq. \ref{eq:Gamma_sb}.
		Hence, this indicator provides information about how much better fuzzy Q-learning agents given $\Gamma$ perform than fuzzy Q-learning agents given $\Gamma^{sb}$. 

		In order to find out, if there is a significant difference between $\Pi_{HQ}^{\Gamma}$ and $\Pi_{HQ}^{baseline}$, Welch's t-tests and Wilcoxon rank-sum tests are applied.\footnote{
		\cites{welch1947generalization} t-test and Wilcoxon rank-sum test \citep{mann1947test} are widely common statistical hypothesis tests.
		In the default case, the first one tests the equality of the means for two independent normally distributed samples with unequal and unknown variances, while the second one, a nonparametric test, tests the null hypothesis that the distributions of two independent samples differ by a location shift of zero or, roughly spoken, that they have the same distribution.} 
		Please be aware that the p-values of hypothesis tests go quickly to zero when very large samples ($10,\hspace{-0.6mm}000$ observations and more) are evaluated, the null hypotheses become statistically significant because the standard errors become extremely small \citep{lin2013research}. 
		To mitigate this phenomenon because, in the simulation study, $10,\hspace{-0.6mm}000$ simulation runs are carried out and, additionally, in each simulation run, the last $100$ observations are recorded for evaluating, one-tailed Welch's t-tests and Wilcoxon rank-sum tests are used with a positive hypothesized mean difference (hereafter abbreviated to $d_H$). 
		Concretely, the null hypotheses are given by $mean(\Pi_{HQ}^{\Gamma})-mean(\Pi_{HQ}^{baseline}) \geq d_H$.
		Note that the Welch's t-test is designed for normally distributed samples but, due to the central limit theorem, it can be assumed that this also holds for the headquarters' profits which are approximately normally distributed. 

		In order to find out whether the headquarters' profits differ significantly in terms of their means (see Welch's t-tests) or their distribution (see Wilcoxon rank-sum tests), the hypothesized mean difference $d_H$ is set to $\Pi_{HQ}^{baseline} / 100$, which can be interpreted as $1\%$ of the headquarters' profit achieved in the baseline scenario. 
		Hence, whenever the t-test result is statistically significant, fuzzy Q-learning agents in the $\Gamma$ scenario perform at least $1\%$ better than fuzzy Q-learning agents in the baseline scenario given a standard significance level of $0.05$. 
		To distinguish the test results, the relative changes are colored as follows: If the p-value of the t-test (rank-sum test) is greater than $0.05$, the cell turns yellow (blue). 
		If both are statistically significant, the cell is green; otherwise the cell is white. 
		
		
		According to Fig. \ref{fig:Plot_05}, non-myopic fuzzy Q-learning agents reach the highest profits, but only as long as they have equal bargaining power. 
		As soon as the distribution of bargaining power changes, the headquarters' profit decreases, whereby $\Pi_{HQ}$ declines faster for non-myopic agents because they react more strongly to a non-symmetrical surplus sharing rule than myopic agents do. 
		Further, the efficiency of the fuzzy Q-learning agents can be seen in the bottom contour plots. 
		While the first-best performance indicator points out that the profit-effectiveness of the headquarters is quite high, especially on the green contours (cf. the bottom left subplot in Fig. \ref{fig:Plot_05}), the second-best performance indicator reflects that, in most parameter constellations, the decision-making behavior of the fuzzy Q-learning agents is much better than the predicted behavior of fully rational utility maximizers (cf. the bottom right subplot in Fig. \ref{fig:Plot_05}). 
		Finally, tests for statistical significance are carried out for each parameter constellation (see the top right subplot in Fig. \ref{fig:Plot_05}). 
		Based on the test results, it can be concluded that, for symmetric marginal cost parameters, a deviation from the theory of subgame perfection equilibrium leads to worse results. 
		In this case, it is more effective to follow the recommendations based on the transfer pricing literature, e.g., \cite{ewert2014interne}, \cite{wagner2008corporate}, \cite{gox2006economic}, and give both divisions equal bargaining power. 
		
		Things look different in scenarios with non-symmetric marginal cost parameters. 
		According to the simulation outputs for $\lambda_S=0.83$ (cf. top right subplot in Fig. \ref{fig:Plot_06}), it is obtained that, for myopic agents with a foresight between $0$ and $0.4$, the surplus sharing parameter should also be set to $\Gamma^{sb}$, but, for non-myopic fuzzy Q-learning agents with $\gamma \geq 0.5$, a deviation from the recommendations leads to better results. 
		Overall, there are $16$ ($14$) scenarios in which the Welch's t-test (Wilcoxon rank-sum test) indicates a significant difference between $\Pi_{HQ}^{\Gamma}$ and $\Pi_{HQ}^{baseline}$. 
		In order to get a reference value for the distribution ratio of the bargaining power for non-myopic fuzzy Q-learning agents, the weighted arithmetic mean of $\Gamma$ is simply calculated, 
		i.e., $\sum_{\Gamma=0.1}^{0.9} \sum_{\gamma=0}^{0.9} \Gamma \cdot BPI[\Gamma,\gamma] / \sum_{\Gamma=0.1}^{0.9} \sum_{\gamma=0}^{0.9} BPI[\Gamma,\gamma]$, where the baseline performance indicator $BPI$ is significant. 
 		The weighted arithmetic mean of $\Gamma$ is $0.33$, which can be considered as a good choice for fuzzy Q-learning agents with a foresight between $0.5$ and $0.9$. 
 		Accordingly, the supplying division receives one third of the bargaining power from the headquarters, while the buying division gets two thirds. 
		This rule of thumb can be particularly relevant because, in general, the headquarters cannot optimally determine the bargaining power, as, e.g., information required to solve the three-stage decision problem is often missing. 

	\subsection{Sensitivity analysis on the variability of state variables and implementation of exploration policies}

	Finally, an extensive sensitivity analysis is performed to check whether the simulation results are robust in terms of the variability of the state variables and the implementation of other exploration policies. 
	In order to evaluate the robustness of the analysis, the environment parameters are adapted so that stochastic fluctuations can now occur. 
	For this purpose, the standard deviations of the state variables $\sigma[\theta_S]$ and $\sigma[\theta_B]$ are set from $0$ to $5$ and $10$, respectively. 
	In addition to the Boltzmann exploration policy, the $\epsilon$-greedy and the upper confidence bound exploration policy, which are introduced in Sec. \ref{sec:Parameter_Settings_and_Simulation_Setup}, are applied. 
	Recall that the $\epsilon$-greedy exploration policy is given in a natural manner and, therefore, can be seen as a benchmark for other more sophisticated exploration policies (\cite{tijsma2016comparing} compare $\epsilon$-greedy, Boltzmann, and UCB policy and evaluate their performance for Q-learning agents). 
	
	Fig. \ref{fig:Plot_07}-\ref{fig:Plot_16} in the appendix show the results from the sensitivity analysis. 
	The contour plots depict the absolute performance of the headquarters' profit resulting from the simulation. 
	Note, while the means of $\Pi_{HQ}$ vary from scenario to scenario, the scale of measure is not set uniformly for all scenarios. 
	This has the effect that the decision-making behavior of the fuzzy Q-learning agents can be better derived, since the headquarters' profit depends positively on the volatility of the markets (see Eq. \ref{eq:optimal_Profit_HQ} and \ref{eq:second_best_HQ}). 
	For example, if $\sigma[\theta_S]=\sigma[\theta_B]=5$, then the additional profit is $(5^2+5^2)/(2\cdot 12)=2.08$ and, for $\sigma[\theta_S]=\sigma[\theta_B]=10$, the headquarters benefits from the volatility of the markets by a value of $8.33$. 
	
	 In the case of symmetric marginal cost parameters (cf. Fig. \ref{fig:Plot_07}), all three exploration policies show a similar picture, whereby fuzzy Q-learning using the $\epsilon$-greedy exploration policy performing worst and fuzzy Q-learning using the Boltzmann exploration policy performing best. 
	 When comparing the columns in Fig. \ref{fig:Plot_07}, fuzzy Q-learning agents exhibit almost a similar decision-making behavior, i.e., they are insensitive to environmental fluctuations. 
	 According to Fig. \ref{fig:Plot_08}, there is no parameter constellation that leads to a better performance than in the baseline scenario with $\Gamma=0.5$. 
	 
	 In the case of non-symmetric marginal cost parameters (cf. Fig. \ref{fig:Plot_09}-\ref{fig:Plot_16}), the learning behavior of the fuzzy Q-learning agents also seems to be robust against the volatilities of the markets. 
	 More importantly, the simulation outputs for $\lambda_S>0.5$ indicate that following the recommendations for setting the surplus sharing parameter $\Gamma$ according to Eq. \ref{eq:Gamma_sb} does not always lead to the best result. 
	 Especially for non-myopic fuzzy Q-learning agents using $\epsilon$-greedy, UCB or Boltzmann exploration policy, the headquarters obtains higher profits when the seller is assigned a greater bargaining power. 
	 Thus, the higher the marginal cost parameter $\lambda_S$, the better it is for all parties involved that the seller gets a higher bargaining power. 
	 In such a case, the headquarters should assign the supplying division a bargaining power based on the calculation of the weighted arithmetic mean. 
	It is also interesting to observe that myopic fuzzy Q-learning agents using a UCB policy contribute to higher profits, if the headquarters assigns the supplying division a lower bargaining power. 




\section{Summary and conclusive remarks}
\label{sec:Conclusion}

	One main reason for studies in the field of specific investments under transfer pricing is to find conditions which provide first-best solutions for the hold-up problem. 
	However, the transfer pricing problem has been developed in the context where divisions are fully individual rational utility maximizers. 
	The underlying assumptions are rather heroic and, in particular, how divisions process information under uncertainty, do not perfectly match with human decision-making behavior. 
	Therefore, this paper relaxes key assumptions and studies whether cognitively bounded agents achieve the same results as fully rational utility maximizers and, in particular, whether the recommendations on managerial-compensation arrangements and bargaining infrastructures are designed to maximize headquarters' profit in such a setting. 
	For this purpose, an agent-based variant of negotiated transfer pricing with cognitively bounded agents is analyzed. 
	
	This paper limits the research to five parameters: (1) the marginal cost parameter in the divisional investment cost function, (2) the surplus sharing parameter in the divisions' incentive compatibility, (3) the discount factor in the fuzzy Q-learning method, (4) the standard deviation of the state variables in regard to the volatility of the markets, and (5) the exploration policy for the action-selection of investments and, in order to produce new theoretical insights, these parameters are systematically vary in a controlled setting. 
	
	Based on an agent-based simulation with fuzzy Q-learning agents, it is shown that (1) in case of symmetric marginal cost parameters, myopic fuzzy Q-learning agents invest only as much as in the classic hold-up problem (second-best solution), while non-myopic fuzzy Q-learning agents invest optimally (first-best solution). 
	(2) in scenarios with non-symmetric marginal cost parameters, $\lambda_S>\lambda_B$, the supplying division always invests in the value range of the second-best solution, regardless of its foresight, while the performance of the buying division depends crucially on the level of foresight, but first-best solutions are not reached in this setting. 
	(3) the simulation output shows that varying the surplus sharing parameter does not lead to significantly better performance in scenarios with symmetric marginal cost parameters.
	However, in scenarios with non-symmetric marginal cost parameters, the choice of the surplus sharing parameter can lead to a significant improvement in the divisions' performance than in the case of the recommended surplus sharing parameters. 
	Therefore, a deviation from the recommended surplus sharing parameter can increase the headquarters' profit. 
	The higher the degree of divisions' foresight, the higher the performance output, if the headquarters assigns the seller a higher bargaining power than the theory of subgame perfection equilibrium recommends. 
	(4) according to the sensitivity analysis, the simulation results are robust with respect to the variability of the state variables and the implementation of $\epsilon$-greedy and upper confidence bound exploration policy. 
	
	The simulation outcomes show that learning by fuzzy Q-learning can be beneficial for economic agents, but whether fuzzy Q-learning matches perfectly with human decision-making behavior is, of course, an empirical question. 
	Beyond that, the findings are based on an agent-based simulation and, therefore, the individual learning behavior of the fuzzy Q-learning agents depends in particular on how the discount factor and the exploration policy are determined. 
	Nonetheless, learning methods from the field of reinforcement learning, such as fuzzy Q-learning, could provide a further clue as to why decentralized firms have the intuition to assign each division equal bargaining power, even if they have different marginal costs parameters. 
	
	Finally, future research could examine the following points: (1) the research focuses on a decentralized firm with only two competing divisions. 
	However, it is also interesting to investigate other decentralized firm settings with more than two divisions, since analytical results are generally not available for such settings. 
	(2) the use of more complicated membership functions, such as those implemented with Gaussian or S-shapes. 
	Such are very popular in the fuzzy logic literature and have properties, such as being differentiable everywhere, but they are more computationally expensive than triangular membership functions.
	(3) future research could also examine a different sequence of events. 
	It would be interesting to know to what extent the simulation results change, if, for example, the state variables are not observed until after the negotiation, or that the investments are also subject to stochastic fluctuations. 

\section*{Appendix: Sensitivity analysis conducted according to scenario 4}
	\vspace{-5mm}
	\begin{figure}[h!]
		\centering
		\includegraphics[width=0.99\textwidth]{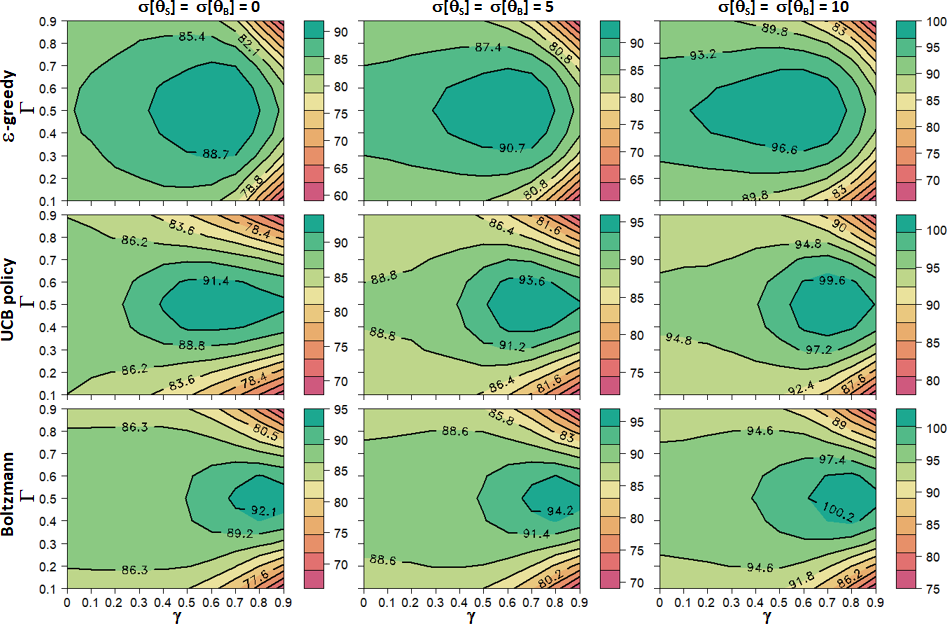}
		\caption{Absolute performance of the headquarters' profit of different exploration policies and standard deviations of state variables given $\lambda_S=\lambda_B=0.5$.}
		\label{fig:Plot_07}
	\end{figure}
	\vspace{-5mm}
	\begin{figure}[h!]
		\centering
		\includegraphics[width=0.99\textwidth]{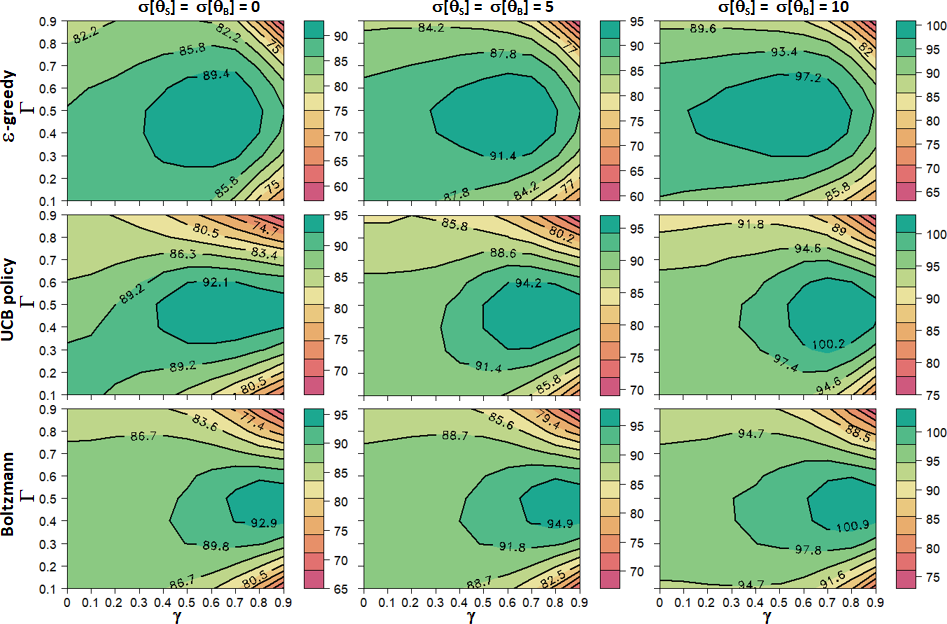}
		\caption{Absolute performance of the headquarters' profit of different exploration policies and standard deviations of state variables given $\lambda_S=0.58$ and $\lambda_B=0.42$.}
		\label{fig:Plot_09}
	\end{figure}

\newpage	
	\begin{figure}[h!]
		\centering
		\includegraphics[width=0.99\textwidth]{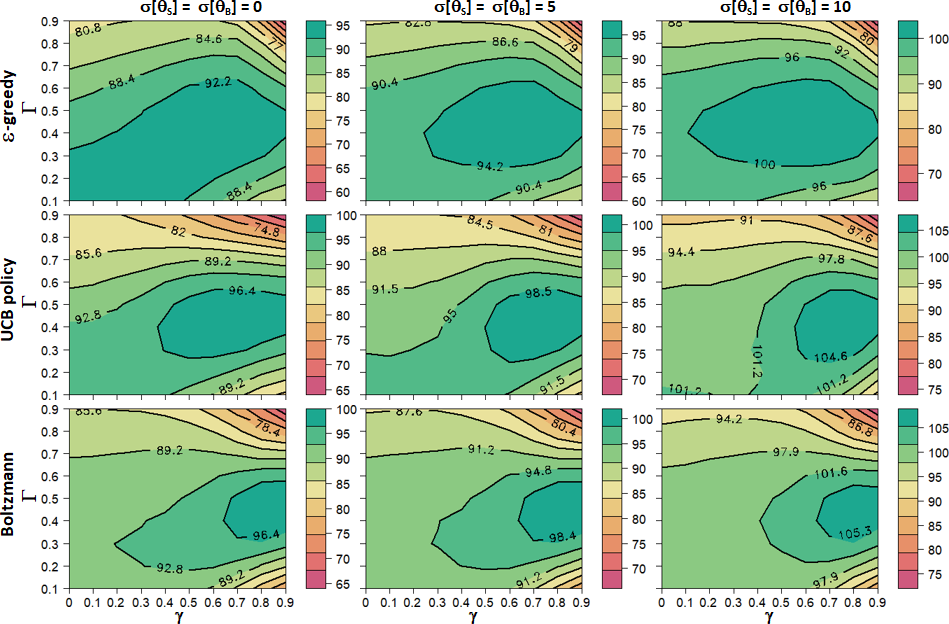}
		\caption{Absolute performance of the headquarters' profit of different exploration policies and standard deviations of state variables given $\lambda_S=0.66$ and $\lambda_B=0.34$.}
		\label{fig:Plot_11}
	\end{figure}
	\begin{figure}[h!]
		\centering
		\includegraphics[width=0.99\textwidth]{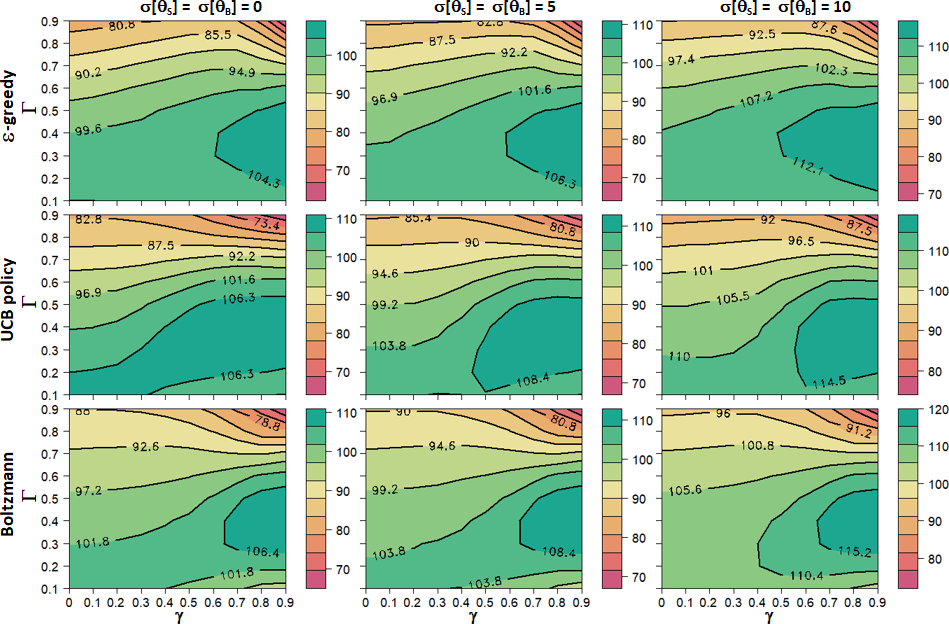}
		\caption{Absolute performance of the headquarters' profit of different exploration policies and standard deviations of state variables given $\lambda_S=0.75$ and $\lambda_B=0.25$.}
		\label{fig:Plot_13}
	\end{figure}

\newpage
	\begin{figure}[h!]
		\centering
		\includegraphics[width=0.99\textwidth]{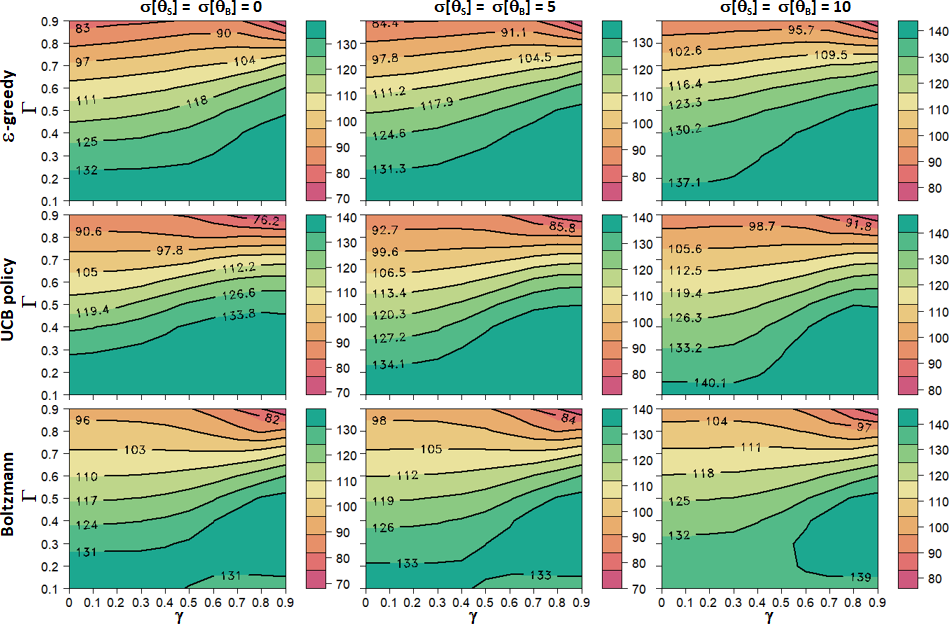}
		\caption{Absolute performance of the headquarters' profit of different exploration policies and standard deviations of state variables given $\lambda_S=0.83$ and $\lambda_B=0.17$.}
		\label{fig:Plot_15}
	\end{figure}
	\begin{figure}[h!]
		\centering
		\includegraphics[width=0.95\textwidth]{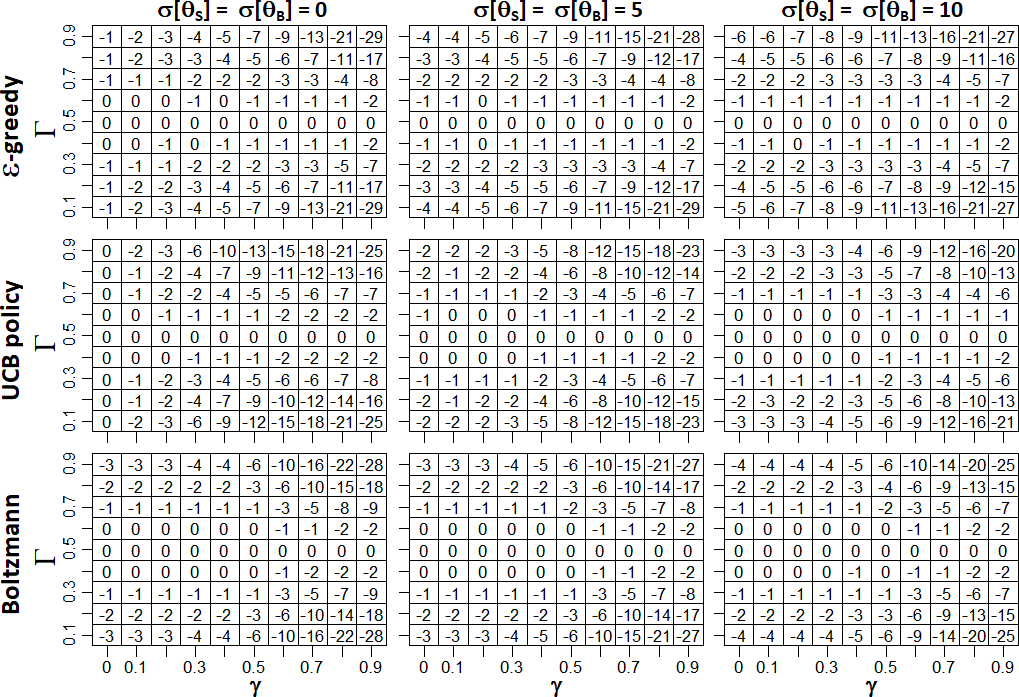}
		\caption{Results of the baseline performance indicator for $\Gamma^{sb}=0.5$ of different exploration policies and standard deviations of state variables given $\lambda_S=\lambda_B=0.5$.}
		\label{fig:Plot_08}
	\end{figure}

\newpage	
	\begin{figure}[h!]
		\centering
		\includegraphics[width=0.96\textwidth]{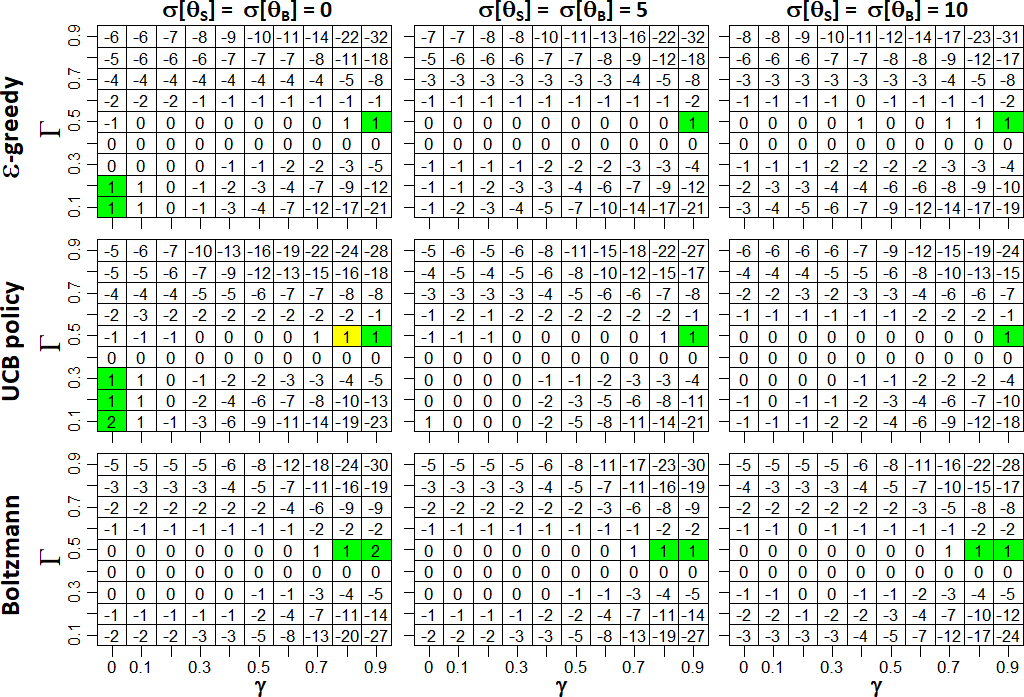}
		\caption{Results of the baseline performance indicator for $\Gamma^{sb}=0.4$ of different exploration policies and standard deviations of state variables given $\lambda_S=0.58$ and $\lambda_B=0.42$.}
		\label{fig:Plot_10}
	\end{figure}	
	\begin{figure}[h!]
		\centering
		\includegraphics[width=0.96\textwidth]{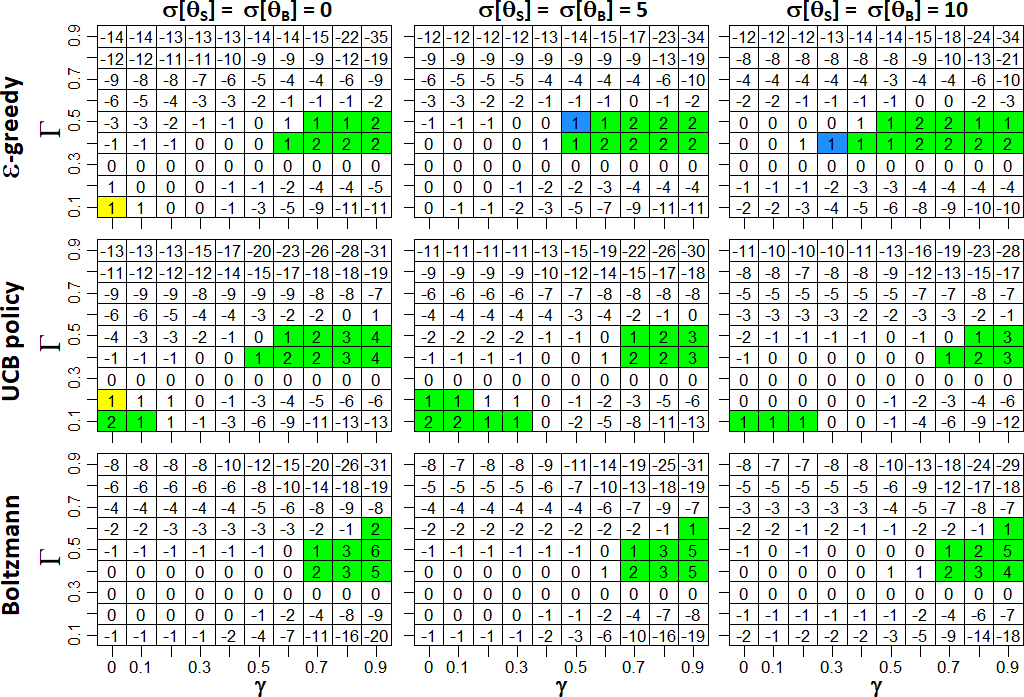}
		\caption{Results of the baseline performance indicator for $\Gamma^{sb}=0.3$ of different exploration policies and standard deviations of state variables given $\lambda_S=0.66$ and $\lambda_B=0.34$.}
		\label{fig:Plot_12}
	\end{figure}

\newpage	
	\begin{figure}[h!]
		\centering
		\includegraphics[width=0.96\textwidth]{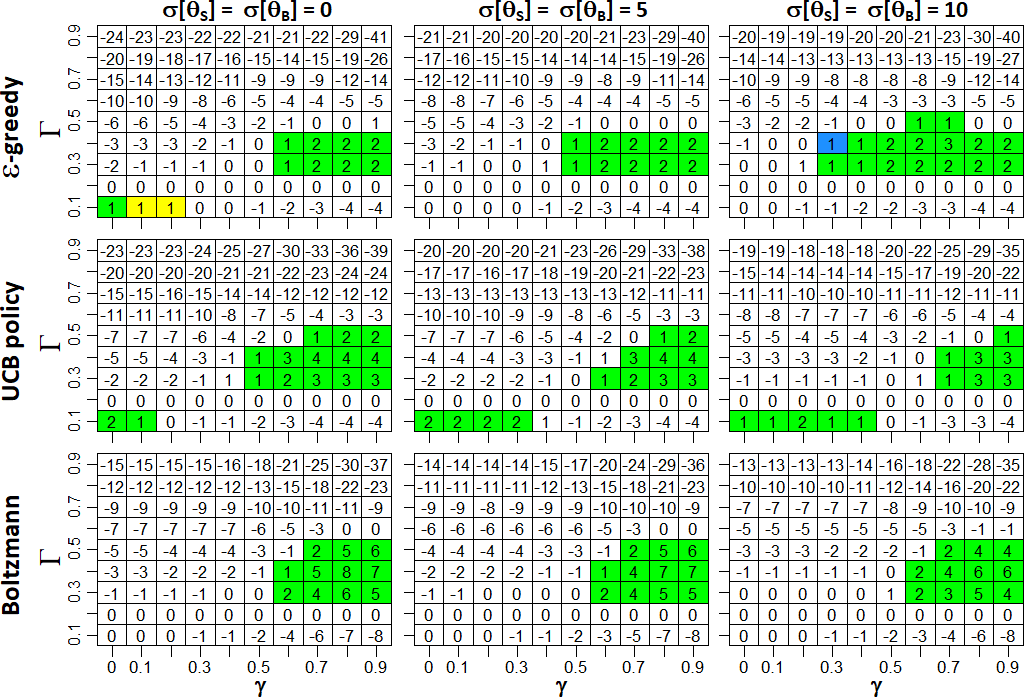}
		\caption{Results of the baseline performance indicator for $\Gamma^{sb}=0.2$ of different exploration policies and standard deviations of state variables given $\lambda_S=0.75$ and $\lambda_B=0.25$.}
		\label{fig:Plot_14}
	\end{figure}	
	\begin{figure}[h!]
		\centering
		\includegraphics[width=0.96\textwidth]{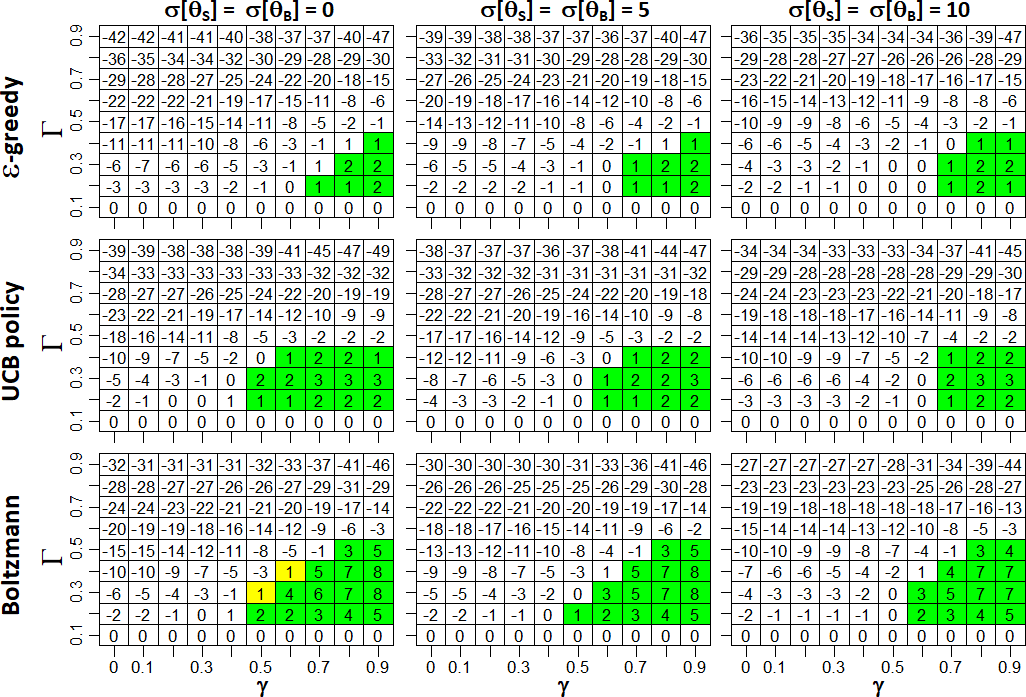}
		\caption{Results of the baseline performance indicator for $\Gamma^{sb}=0.1$ of different exploration policies and standard deviations of state variables given $\lambda_S=0.83$ and $\lambda_B=0.17$.}
		\label{fig:Plot_16}
	\end{figure}

\newpage
\bibliographystyle{spbasic}      
\bibliography{biblio}   

%
%

\end{document}